\documentclass[aps,prb,twocolumn,amsmath,amssymb,floatfix]{revtex4-1}
\usepackage{tabularx}
\usepackage{bm}
\usepackage{euscript}
\usepackage{graphicx}
\usepackage{color}
\usepackage{amsfonts}
\usepackage{exscale}
\usepackage{amsbsy}
\usepackage{textcomp}
\usepackage{bbold}
\usepackage[caption=false]{subfig}
\usepackage{hyperref,placeins}

\newcommand{\be}{\begin{eqnarray}}
\newcommand{\ee}{\end{eqnarray}}

\def\cdag{\ensuremath c^{\dag}}
\def\c{\ensuremath c^{}}

\def\k{\ensuremath \bm{k}}
\def\q{\ensuremath \bm{q}}

\def\kp{\ensuremath \bm{k^\prime}}
\def\kpp{\ensuremath k^\prime}
\def\Q{\ensuremath \bm{Q}}
\def\psibar{\bar{\psi}}
\def\Lambday{\ensuremath \Lambda_y}
\def\Lambdax{\ensuremath \Lambda_x}

\begin{document}

\title{Crisscrossed stripe order from interlayer tunneling in hole-doped cuprates}
\author{Akash V. Maharaj,$^1$ Pavan Hosur,$^1$ and S. Raghu$^{1,2}$}
\affiliation{$^{1}$Department of Physics, Stanford University, Stanford, California 94305, USA}
\affiliation{${}^{2}$SLAC National Accelerator Laboratory, Menlo Park, CA 94025, USA}

\date{\today}

\begin{abstract}

 Motivated by recent observations of charge order in the pseudogap regime of hole-doped cuprates, we show that  {\it crisscrossed} stripe order can be stabilized by coherent, momentum-dependent interlayer tunneling, which is known to be present in several cuprate materials. We further describe how subtle variations in the couplings between layers can lead to a variety of stripe ordering arrangements, and discuss the implications of our results for recent experiments in underdoped cuprates.  
\end{abstract}

\maketitle

\section{Introduction}
There is a growing body of experimental evidence suggesting that charge order develops within the pseudogap regime, and competes with superconductivity in several of the hole-doped cuprates.  
In low magnetic fields, the range of doping over which Fermi surface reconstruction has been inferred from transport measurements in YBa$_2$Cu$_3$O$_{6+x}$ (YBCO)\cite{Chang2010} coincides roughly with the onset of short ranged, incommensurate charge order as seen in X-ray scattering measurements.\cite{XrayYBCO, chang2012direct,RexsYBCO}  NMR measurements of quadrupolar frequency broadening have also revealed that this low field order is static,\cite{wu2014short} however there is also NMR evidence for unidirectional charge order induced at high magnetic fields,\cite{NMRYBCO} in the same region where quantum oscillation measurements find small electron-like pockets.\cite{Proust2007,Proust2007a,Sebastian2009,Sebastian2010,sebastian2012towards} While the relationship between the low and high field ordering tendencies remains uncertain, these experiments clearly reveal that charge order is an important feature of cuprate physics.

Inelastic X-ray scattering experiments have shown that charge order in underdoped YBCO is short ranged, and exhibits peaks at two wave-vectors, $\Q_x = 2\pi(q_x, 0, 0.5)$ and $\Q_y = 2\pi(0, q_y, 0.5)$, where $q_x \approx q_y$ is weakly doping dependent,\cite{huecker2014competing,blancoXray} and incommensurate with the underlying lattice ($q_{x,y} \approx 0.31$ for 12$\%$ hole doped samples\cite{XrayYBCO,chang2012direct}). Furthermore, while the X-ray peak magnitudes at $\Q_x$ and $\Q_y$ are roughly equal near one-eighth doping, significant anisotropies develop at other hole concentrations.\cite{blanco2013momentum,blancoXray} A common interpretation of these experiments is that they reveal a form of checkerboard charge order,\textit{ i.e.} each layer consists of equal amplitudes of density waves ordered at $\Q_x$ and $\Q_y$, but it is also possible that the anisotropy in peak magnitudes at some dopings reflects an underlying tendency to unidirectional or `striped' states. Striped phases, which have been observed in doped LaBaCuO$_4$,\cite{Tranquada2004} LaSrCuO$_4$,\cite{suzuki1998observation} and Bi$_2$Sr$_2$CaCu$_2$O$_8$,\cite{howald2003periodic} break distinct symmetries compared to checkerboard order, and their presence in YBCO would suggest that the physics of the copper oxide plane is universal. The X-ray patterns observed in YBCO could correspond to domains of $x$- and $y$-directed stripes,\footnote{There are many subtle issues in distinguishing striped from checkerboard order from X-ray data when the order is not long ranged. See Sec.V of the Supplemental Material of \citep{nie2013quenched} also Ref.~\onlinecite{robertson2006distinguishing}} but a more exotic possibility is that the density wave order is staggered, orienting along $Q_x$ in one layer then along $Q_y$ in adjacent layers. We refer to such a state as a `crisscrossed pattern of stripes.' 

A crisscrossed pattern is reminiscent of the stripe order found in the `214' family of cuprates. In one-eighth doped LaBaCuO$_4$ for instance, a pattern of orthogonal stripes has been confirmed by X-ray and neutron diffraction studies,\cite{Tranquada2004,Tranquada2009} and is commonly understood as being induced by the alternating tilt directions of CuO${}_6$ octahedra from one layer to the next. While it is appealing to speculate that the  stripe ordering tendencies might be universal among different cuprates, it is unclear why a crisscrossed pattern would occur in a weakly orthorhombic material such as YBCO, where  such structural `source terms' for orthogonality are absent. 
Here, we identify, at a qualitative level, aspects of the cuprate electronic structure that help stabilize a crisscrossed pattern of stripe order.  

Given the layered nature of the cuprate materials, it is natural to suppose that the dominant correlations - including those responsible for superconductivity, anti-ferromagnetism, charge order, and the pseudogap - occur within a plane of the material.  However, in principle, it is conceivable that weak couplings {\it between} layers can lead to new arrangements of these phases, an example being crisscrossed patterns of stripes.  Our main conclusion is that in the presence of momentum-dependent interlayer tunneling - precisely of the form that has been shown to occur in many cuprate materials\cite{chakravarty1993interlayer,OkAndersen} - phase transitions can occur between parallel and crisscrossed patterns of stripe order.   To the extent that the physics within a  copper oxide plane is universal, it is plausible that subtle changes in the couplings between the layers can account for the variations in density-wave correlations observed at different dopings of YBCO.

Our results are based on a mean-field analysis of a bilayer system, in which there are strong tendencies to form unidirectional charge density waves (CDWs) in each layer. We do not speculate on the microscopic origins of the density waves here; this question has been addressed to a considerable extent in the literature.\cite{zaanen1989charged,machida1989magnetism,schulz1990incommensurate,emery1993frustrated,emery1999stripe,white1998density,chakravarty2001hidden,sachdev2013bond, wang2014charge,laughlin2014fermi}  Instead, we will {\it assume} that each layer exhibits a propensity towards unidirectional CDW (stripe) formation and ask how the coupling between layers affects the charge order. We first consider the Landau-Ginzburg theory of a bilayer system in Sec.~\ref{sec:Landau}, where a variety of symmetry-distinct stripe orderings can arise from simple phenomenological interactions. In Sec.~\ref{sec:microscopic} we perform an explicit Hartree-Fock study of the bilayer system at zero temperature, which demonstrates the contrasting roles of interlayer tunneling and Coulomb interactions in controlling the relative orientation of stripes.  Next, we study the system at finite temperature in Sec.~\ref{sec:finiteT}, while in in Sec.~\ref{sec:discussion} we discuss how our model bilayer results can be generalized to multiple layers in three dimensional systems such as YBCO. We close with a discussion of how our results can be interpreted in the context of the cuprates, emphasizing the doping dependence of charge order, as well as the role of quenched disorder.

%----------------------------------------------------------------------------------------------------------------------------------------------------

\section{Landau Ginzburg theory of crisscrossed stripes}\label{sec:Landau}
The various arrangements of striped phases and the symmetries they break, can be understood by first considering the Landau Ginzburg (LG) theory of a layered disorder-free tetragonal system in which incommensurate, unidirectional charge density waves (stripes) onset continuously at a charge ordering temperature $T_{\text{CO}}$.\footnote{The existence of various types of density wave order, both at zero and finite temperature in the single layer model has been previously studied in the presence of incommensuration\cite{DelMaestro1,DelMaestro2}}  The $C_4$ symmetry of the lattice allows for density waves in two inequivalent directions. We therefore introduce a complex, vector order parameter $ \vec{\phi}_{\lambda} = (\phi_{\lambda,x},\phi_{\lambda,y})$ to represent the charge modulations in layer $\lambda$, such that the charge density at position $\vec{r}$ is given by
\begin{align}
\rho_{\lambda}(\vec{r}) = \bar{\rho} + \left[ \phi_{\lambda,x}(\vec{r})e^{iQ x} + \phi_{\lambda,y}(\vec{r})e^{iQ y} + c.c\right].
\end{align}
The ordering vector magnitude $Q$ is equivalent in the $x$ and $y$ directions and $\bar{\rho}$ is the average charge density. 
Near to $T_{\text{CO}}$, the free energy of a layered system can be expanded in powers of these order parameters. It is natural to divide the free energy into two parts, $
F = F_{\text{in-plane}} + F_{\text{inter-plane}},
$
where the in-plane contribution contains the dominant energy scales, is the same for all layers and has the general form
\begin{align}
F_{\text{in-plane}} &=r\left( |\phi_{\lambda,x}|^2 + |\phi_{\lambda,y}|^2\right) +  u\left( |\phi_{\lambda,x}|^2 + |\phi_{\lambda,y}|^2\right)^2 \nonumber
\\&\,\,\,+ w|\phi_{\lambda,x}|^2|\phi_{\lambda,y}|^2,
\end{align}
to quartic order in $\phi$'s. The mechanism that drives charge ordering is contained in this expression, with the sign of $w$ determining whether there is striped ($w>0$) or checkerboard order ($w<0$). For a single tetragonal layer, only one component of $\vec{\phi}$ has a non-zero expectation value in the striped state, while in a checkerboard state both components of develop \textit{equal} magnitudes. The checkerboard phase therefore breaks a $U(1)\times U(1)$ symmetry corresponding to the phases of the incommensurate density waves in the $x$ and $y$ directions. On the other hand, the striped phase breaks a single $U(1)$ symmetry in addition to a discrete Ising symmetry associated with its direction. We note that the above distinctions between checkerboard and stripe order are not present in an orthorhombic environment where there can be a split transition, below which both components of $\vec{\phi}_{\lambda}$ are non zero but their relative amplitudes controlled by the degree of orthorhombicity.

Our microscopic calculations of Sec.~\ref{sec:microscopic} find predominantly stripe order in single layers, so we henceforth assume $w>0$ when generalizing to bilayers. The relative orientation of stripes in neighboring layers is then determined by the $F_{\text{inter-plane}}$. In terms of $\vec{\phi}_{\lambda}$ for layers $\lambda = 1,2$ of a tetragonal bilayer system, this free energy expansion takes the form
\begin{align}
F_{\text{inter-plane}} &= \alpha\left(\vec{\phi}_{1}\cdot\vec{\phi}^{*}_{2} + c.c.\right)+ \beta\left\lvert\vec{\phi}_{1}\cdot\vec{\phi}^{*}_{2}\right\rvert^2 + \gamma\left\lvert \vec{\phi}_{1} \times \vec{\phi}_{2}\right\rvert^2.
\label{eq:LGinterlayer}
\end{align}
Inter-plane couplings therefore have two main effects: near $T_{\text{CO}}$ where the order parameter magnitude is small, the bilinear term dominates, favoring parallel stripes with the sign of $\alpha$ determining their relative phase. Then, provided $\alpha$ is relatively small, there can be a transition to perpendicularly stacked (crisscrossed) stripes as the temperature is lowered and the magnitude of $\vec{\phi}$ grows so that the biquadratic terms become important. This transition is possible if  $ \beta > \gamma$, while the opposite sign chooses parallel stripes. Intuitively the difference of the two biquadratic terms in Eq. \ref{eq:LGinterlayer} contains a `quadrupolar' interaction, which is repulsive when $\beta > \gamma$, favoring the crisscrossed stripe order.

While the parallel phases break identical symmetries as their single layer striped counterparts, the crisscrossed phase breaks $U(1)\times Z_2$ symmetry in each layer, but preserves 90 degree rotation followed by mirror reflection through a horizontal plane lying half-way between the bilayer. Importantly, the crisscrossed phase breaks translation symmetry in both $x$- and $y$- directions, so the two components of $\vec{\phi}_{\lambda}$ are generically non-zero within a given layer $\lambda$. However the amplitudes of the density wave modulations in the two directions will be strongly anisotropic, reflecting the underlying tendency towards stripe order. As we show in the next sections, an explicit Hartree-Fock minimization finds either parallel or crisscrossed phases with precisely these symmetries, where the coefficients $\alpha, \beta$ and $\gamma$ are controlled by the strength and form of interlayer tunneling and interactions.

%----------------------------------------------------------------------------------------------------------------------------------------------------

\section{Microscopic model}\label{sec:microscopic}
 Our microscopic model involves two layers that are coupled via coherent electron tunneling (of strength $t_{\perp}$) and Coulomb interactions ($V_{\perp}$). As discussed earlier, we impose an effective in-plane interaction to favor CDWs at wavevectors of our choosing. For simplicity, we ignore the electron spin degrees of freedom, and study density waves of commensurability 2 and 3,\footnote{While a period 3 density wave is close to the experimentally observed ordering vector of $|\Q|\approx2\pi\cdot 0.31$, we have chosen low order commensurate density waves for technical convenience as this greatly simplifies the sums over the reduced Brillouin zone (e.g. a period 3 density wave in both $x$ and $y$ directions corresponds to  a Brillouin zone (BZ) that is 1/9th the original BZ. This means that the two layer Hamiltonian is an $18\times18$ matrix, which is relatively simple to numerically analyze).}  allowing ordering in the lattice $x$ and $y$ directions.  Denoting the spinless electron creation operators in layer $\lambda$ by $\cdag_{\k\lambda}$ where $\k$ is an in-plane momentum, the Hamiltonian has the form $H = H_{t} + H_{V}$, with 
\begin{align}
H_{t} &= \sum_{\k;\lambda,\lambda^\prime} \left[\left(\varepsilon(\k)-\mu\right)\delta_{\lambda\lambda^{\prime}} + t_{\perp}(\k)\tau^{x}_{\lambda\lambda^\prime}\right] \left(\cdag_{\k\lambda} \c_{\k\lambda^\prime} + h.c\right) \nonumber\\
H_{V}&= \sum_{\k,\k^\prime,\q;\lambda,\lambda^\prime} \hat{V}_{\lambda\lambda^\prime}(\q) \cdag_{\k+\q\lambda} \c_{\k\lambda} \cdag_{\kp-\q\lambda^\prime} \c_{\kp\lambda^\prime},
\label{eq:mainham} 
\end{align} 
where $\tau^{x}$ is the Pauli $x$-matrix, and the bare band structure is chosen to represent a typical single layer cuprate bandstructure, \textit{i.e.} $\varepsilon_{\k} = -2t(\cos{k_x} + \cos{k_y}) + 4t^{\prime}\cos{k_x}\cos{k_y}$ , with $t=1$ and $t^{\prime} = 0.3t$, while $\mu$ is chosen to work at half filling. Meanwhile the interlayer tunneling $t_{\perp}(\k)$ is
\begin{align}  
t_{\perp}(\k) = \frac{t_{\perp}}{4}\left(\cos{k_x} - \cos{k_y}\right)^2,
\label{eq:interlayerT}
\end{align}
though we will later vary this form to determine how interlayer tunneling can affect our results. This form of interlayer tunneling in the cuprates was originally inferred from the reduced splitting of hybridized bands in nodal directions,\cite{chakravarty1993interlayer} and subsequently confirmed for tetragonal structures by ab-initio studies.\cite{OkAndersen}

For stripes at ordering vectors $\Q_{x} = (Q,0)$ and $\Q_{y}=(0,Q)$, the in-plane interaction must be attractive and peaked at these vectors, so we adopt the model interaction $\hat{V}_{11}(\q) = \hat{V}_{22}(\q) = -V(\delta_{\q,\Q_{x}} + \delta_{\q,\Q_{y}})$, with ordering vectors of magnitude $Q=\pi$ and $Q= 2\pi/3$. Since there is no nesting instability of the Fermi surface to CDW order at these wavevectors, we require in-plane interaction magnitudes of the order of the bandwidth to generate density waves. Our explicit results were obtained for $V$'s in the range $5t$ to $10t$. Finally, we choose the inter-plane interactions to be short ranged, with $\hat{V}_{12}(\q) = \hat{V}_{21}(\q) = V_{\perp}/2$. 

We now introduce a trial Hamiltonian that allows density waves in both the $\bm{\hat{x}}$ and $\bm{\hat{y}}$ directions, and minimize the free energy $F_{0} = F_{tr} + \langle H - H_{tr} \rangle_{tr}$, where expectation values are taken with respect to the trial Hamiltonian. This quadratic trial Hamiltonian has the form $H_{tr} = H_{t} + H_{\phi}$, with  
\begin{align}
H_{\phi} &= \sum_{\k;\lambda} \phi_{\lambda,x}\cdag_{\k+\Q_{x} \lambda} \c_{\k\lambda} +  \phi_{\lambda,y}\cdag_{\k+\Q_{y} \lambda}\c_{\k\lambda} + (h.c.)
\end{align}
where $\phi$'s are now variational parameters, and we also adjust the chemical potential in order to work at fixed density. The free energy minimization yields self consistency equations for $\phi_{\lambda,i}$'s of the form
\begin{equation}
\phi_{\lambda,i} = \sum_{\k} -V\langle \cdag_{\k\lambda}\c_{\k+\Q_{i}\lambda}\rangle_{_{tr}} + V_{\perp}\langle \cdag_{\k\lambda^{\prime}}\c_{\k+\Q_{i}\lambda^{\prime}}\rangle_{_{tr}},
\label{eq:selfconsist}
 \end{equation}
 where $\lambda^{\prime} \ne \lambda$. These equations are numerically iterated to achieve self consistency. 
\begin{figure}
\begin{center}
        \includegraphics[width=0.49\textwidth]{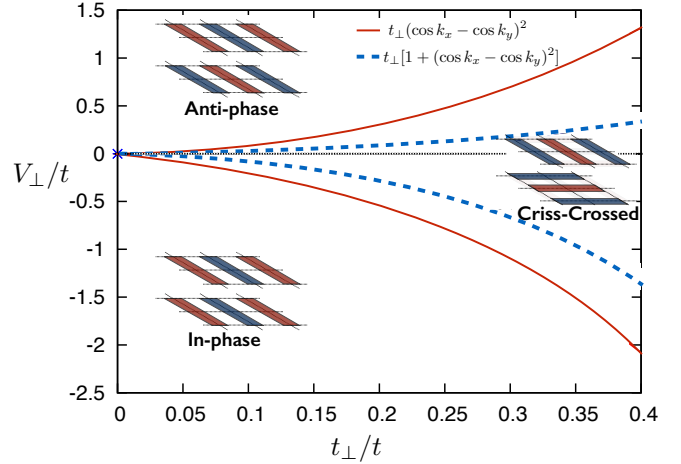}
              \caption{The phase diagram obtained for the microscopic model of Eq.~\ref{eq:mainham} for different forms of interlayer tunneling, with $|\Q|=2\pi/3$ density waves. Adding a $\k$ independent component to the interlayer tunneling shifts the red phase boundaries to the blue dashes, stabilizing the anti-parallel phase. The phase diagram is qualitatively similar for period 2 density waves.}
               \label{fig:phasediag1}
\end{center}
\end{figure}

For a single (uncoupled) layer, we find stripes in either direction as degenerate solutions. We stress that the checkerboard pattern is not the variational minimum of the present model; instead stripe phases have a lower variational free energy. This reflects the lack of a Fermi surface instability at these wavevectors; while checkerboard states are expected in weak coupling studies where they destroy larger portions of the Fermi surface than striped states and hence are energetically favorable, this intuition does not apply when $\phi$'s are large due to stronger couplings. 

Upon coupling the layers, we obtain three classes of self consistent solutions at zero temperature, representing the different orientations the stripes can form. The CDW order parameter in these phases take on the following values (zero if not shown):
\begin{enumerate}
\item {\bf Parallel in-phase}: $\phi_{1,x} = \phi_{2,x} = \Phi$
\item {\bf Parallel anti-phase}: $\phi_{1,x} = -\phi_{2,x} = \Phi$
\item {\bf Crisscrossed}: $\phi_{1,x}=\phi_{2,y}=\Phi$, and\\\qquad\qquad $\phi_{1,y}=\phi_{2,x} =\delta \ll \Phi$,
\end{enumerate}
and equivalent solutions are obtained by interchanging $x$ and $y$. We see that the symmetry considerations of the previous sections are manifest; while the two parallel phases have only a single non-zero component of $\vec{\phi}$ in each layer, both components are non-zero in the crisscrossed phase, but with strongly anisotropic magnitudes. The magnitude of this anisotropy is controlled by the ratio of inter- to intra-layer couplings, and is typically on the order of  $|\delta/\Phi| <0.1$ for all regions of the phase diagram. While all three states are valid self consistent solutions at zero temperature, comparison of their free energies yields a phase diagram in the plane of $V_{\perp}$ and $t_{\perp}$ as shown in Fig.~\ref{fig:phasediag1}.

\begin{figure}
\begin{center}
        \includegraphics[width=0.47\textwidth]{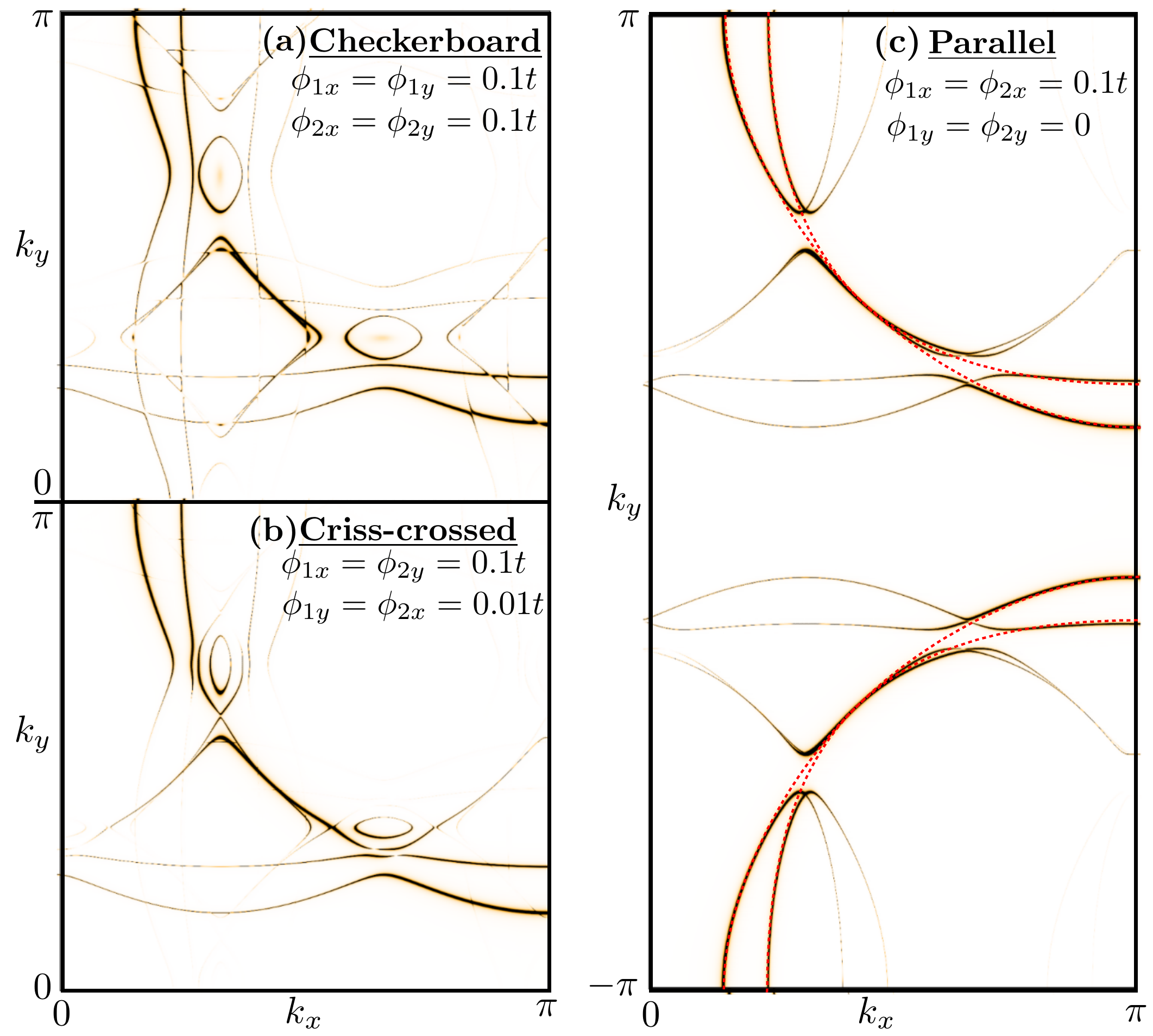}
              \caption{The electron spectral functions of the reconstructed Fermi surfaces of $|\Q| = 2\pi/3$ density waves in bilayer systems with weak $\k$ dependent interlayer tunneling, $t_{\perp} /4= 0.1t$. \textbf{(a)} For reference we show the reconstruction for a checkerboard pattern in each layer i.e. all $\phi$'s are equal in magnitude, \textbf{(b)} a crisscrossed pattern of stripes, and \textbf{(c)} a parallel pattern of stripes. The red dashed lines overlaid in \textbf{(c)} show the bare (unreconstructed) Fermi surface with bilayer splitting ($t_{\perp}/4 = 0.1t$).}
               \label{fig:fermisurfaces}
\end{center}
\end{figure}

We see that parallel striped states are favored when Coulomb interactions are the important interlayer energy scale: for a positive (repulsive) $V_{\perp}$ , the density waves are out of phase as this enables the system to minimize the repulsion. Conversely, an effective attraction between charges supports an in-phase alignment of density waves. On the other hand, when coherent interlayer tunneling is the dominant interlayer energy scale, a crisscrossed phase is stabilized. There are first order phase boundaries with $V_{\perp} \sim t^{2}_{\perp}$ between parallel and crisscrossed phases.

Up to now, all results have been for an interlayer tunneling of the form given in Eq.~\ref{eq:interlayerT}. An interesting feature is the shifting of phase boundaries `downwards'  upon adding some uniform, $\k$ independent component to the interlayer tunneling (blue dashed phase boundaries in Fig.~\ref{fig:phasediag1}). Indeed, if the tunneling is purely uniform, the crisscrossed phase can only be stabilized when interlayer interactions are attractive.  This is because a momentum-dependent interlayer tunneling induces effective interactions (which in the small tunneling regime are $\mathcal O(t_{\perp}^2/t)$) that weaken the Coulomb repulsion between layers; for a momentum-independent tunneling, this suppression is absent.

Finally, let us examine the Fermi surface reconstructions produced by these phases. Figure~\ref{fig:fermisurfaces} shows the electronic spectral function of the reconstructed Fermi surfaces in the parallel and crisscrossed phases, along with a checkerboard state for reference. This has been done for the parent band structure given by $H_t$ in Eq.~\ref{eq:mainham}, where interlayer tunneling results in a bilayer splitting in all but the nodal directions. We show the Fermi surfaces produced by period three density waves of small amplitude, having treated the density wave magnitude as an external parameter in order to illustrate the reconstruction. The parallel and crisscrossed phases produce very different reconstructions reflecting their distinct broken symmetries. The former has a large hole pocket, while the checkerboard and crisscrossed patterns have similar Fermi surfaces, the major features of which are an electron pocket in the nodal direction, with smaller hole pockets nearby.\cite{eun2012quantum}$^{,}$\footnote{A recent pre-print considering quantum oscillations in single layers with incommensurate density wave order found very similar patterns of Fermi surface reconstruction.\cite{allais2014connecting}}  Thus, while the checkerboard and crisscrossed phases are likely to arise from disparate microscopic mechanisms, they each break translation symmetry in both $x$- and $y$-directions and resulting in similar Fermi surface reconstructions. The formation of such nodal electron pockets from checkerboard charge order was first proposed by Harrison and Sebastian,\cite{harrison2011protected,sebastian2012towards} and while several quantum oscillation measurements have found evidence for the presence of small electron pockets,\cite{Proust2007,Proust2007a,Sebastian2009,Sebastian2010} their nodal location has only recently been experimentally deduced.\cite{sebastian2014normal}

%----------------------------------------------------------------------------------------------------------------------------------------------------

\section{Finite Temperature}\label{sec:finiteT}
\begin{figure}[t]
\begin{center}
        \includegraphics[width=0.5\textwidth]{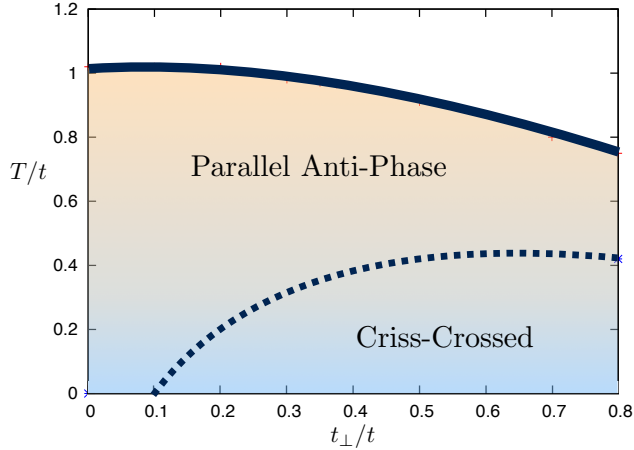}
              \caption{The phase diagram as a function of temperature for a small positive $V_{\perp}=0.1t$, showing a continuous transition from the state with no CDW order into the Parallel Anti-Phase state (solid curve), followed by a first order transition to a crisscrossed configuration of density waves (dashed curve).  }
               \label{fig:phasediag2}
\end{center}
\end{figure}

From the phenomenological considerations of Sec.~\ref{sec:Landau}, we expect a parallel phase to emerge at the transition temperature (where the bilinear term in the free energy is important), with the crisscrossed phase possibly arising at a lower temperature. Numerically solving the self consistency relations (Eq.~\ref{eq:selfconsist}) at finite temperature reveals precisely this behavior. Fig.~\ref{fig:phasediag2} shows the phase diagram as a function of $t_{\perp}$ for a small positive $V_{\perp}$, where the upper transition is continuous, while the crisscrossed phase onsets at a first order transition. 

We may further verify these results by deriving the Landau Ginzburg theory corresponding to our microscopic model. Since the charge density waves onset continuously, we integrate out fermions in the model of Eq.~\ref{eq:mainham} and expand the free energy in powers of the order parameter fields near to this transition.  Once more, our interest lies in the interlayer terms, which take the form shown in Eq.~\ref{eq:LGinterlayer}. The details of this procedure are given in the Appendices, where we evaluate the coefficients $\alpha$, $\beta$ and $\gamma$ of the interlayer free energy (given explicitly by Equations~\ref{eq:alpha1}-\ref{eq:gamma1}), employing the usual hotspot approximations\cite{wang2014charge} in evaluating these Fermi surface integrals. The coefficient $\alpha$ has contributions from the both the interlayer interactions  at $\mathcal{O}(V_{\perp}/V^2)$, and interlayer tunneling at $\mathcal{O}(t^{2}_{\perp}/t^{2})$. On the other hand, the quartic coefficients depend purely on interlayer tunneling with lowest order contributions of $\mathcal{O}(t^{2}_{\perp}/t^{2})$ from several hotspots.

The main result here is that due to the suppression of tunneling near to the hotspots of the Fermi surface (brought about by the momentum dependent interlayer tunneling), the resulting induced Coulomb interaction governed by the coefficient $\alpha$ is strongly and preferentially suppressed (compared to $\beta$ and $\gamma$). This therefore reduces the tendency of forming parallel states and is the mechanism allowing a transition to the crisscrossed phase as the temperature is lowered and the quartic terms to dominate.  It is important to note that while our numerical studies were done for commensurate CDW's, these phenomenological  arguments hold for general wavevectors and so are appropriate for the experimentally relevant incommensurate charge orders seen in the cuprates.

%----------------------------------------------------------------------------------------------------------------------------------------------------

\section{Discussion}\label{sec:discussion}
At this stage, our studies of bilayer systems have revealed the presence of perpendicular or parallel striped phases depending on the relative importance of tunneling and interactions between layers. It is then natural to ask how these states are generalized in a layered, three dimensional structure. Focussing on a YBCO-like structure, with an inter-bilayer distance that is roughly three times the intra-bilayer separation, we can generally expect the tunneling within a bilayer to be much stronger than between separate bilayers. It is therefore likely that neighboring layers in adjacent bilayers will be in the Coulomb dominated regime of Fig.~\ref{fig:phasediag1}, forcing the stripes in adjacent layers but neighboring bilayers to be parallel but out of phase.

\begin{figure}
\begin{center}
        \includegraphics[width=0.45\textwidth]{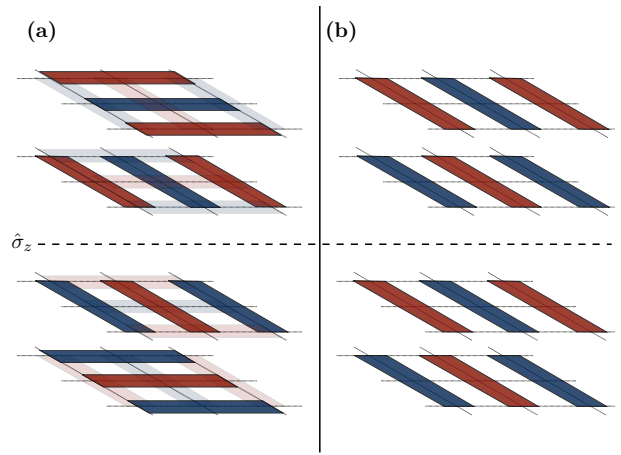}
              \caption{The multilayer states that result from generalizing \textbf{(a)} crisscrossed phases and \textbf{(b)} parallel phases. Red (Blue) represent regions of high (low) charge density, and the lighter colored stripes in \textbf{(a)} reflect the small transverse modulations of magnitude $\delta$ within each layer for the crisscrossed phase. Both break horizontal mirror plane symmetries ($\hat{\sigma}_z$), while vertical translation symmetry is broken by the state in \textbf{(a)}.}               \label{fig:stripe_arrange}
\end{center}
\end{figure}

This results in two possible multi-layer configurations as illustrated in Fig.~\ref{fig:stripe_arrange}. Where the density waves within a bilayer are in the crisscrossed phase, the multi-layer generalization breaks (discrete) translation symmetry in the $c$ directions, along with horizontal mirror symmetries. In addition, vertical mirror plane symmetries in both directions may be broken when incommensurate density waves---which are not pinned to a lattice---are present (these incommensurate striped phases are not shown in Fig.~\ref{fig:stripe_arrange}). Alternatively, when density waves are in the parallel phase within each bilayer, translation symmetry in the $c$ direction is preserved, although the same horizontal mirror symmetry is broken. Here, only a single vertical mirror plane symmetry is broken when incommensurate stripes are present.

So far, we have not taken into account the role of quenched disorder, which likely plays a deciding role in the cuprate phase diagram.  Incommensurate charge density waves cannot exhibit long range order in three spatial dimensions in the presence of random field disorder; nevertheless, the component of the charge order corresponding to discrete symmetry breaking can retain long range correlations.\cite{ImryMa}  The static nature of the charge order,\cite{wu2014short} along with evidence for long ranged nematicity\cite{lawler2010intra} supports the notion that disorder is important to cuprate phenomenology, and was explored recently in Ref. \onlinecite{nie2013quenched}.  In the present context, the parallel stripe phase has a nematic component which can remain long range ordered in the presence of quenched random fields.  In the case of criscrossed stripes, the order breaks mirror symmetries, which remain stable to weak random field disorder.\footnote{Under certain conditions, the crisscrossed phase may become chiral with the handedness of the chirality breaking a $Z_2$ symmetry. This is possible when bond density waves (with order parameter $\vec{\bm{\Delta}}_{\lambda}$) are also present in the clean system. A pattern of commensurate charge and bond density waves, which is perpendicular from layer to the next, but in which the CDW switches sign between alternate layers will be chiral, with the order parameter $\chi = \sum_{\lambda}\bm{\hat{z}}\cdot\left(\bm{\hat{n}}_{\lambda}\times \bm{\hat{n}}_{\lambda+1}\right)$, where the vector $\bm{n}_{\lambda}$ in each layer is $\bm{n}_{\lambda} =  \phi_{\lambda,x} \Delta^{*}_{\lambda,x}\bm{\hat{x}} + \phi_{\lambda,y}\Delta^{*}_{\lambda,y}\bm{\hat{y}}$. This suggests a purely electronic mechanism for establishment of a chiral phase, unlike previous studies.\cite{jasper1,jasper2,jasper3}}  The occurrence of a zero-field Nernst effect, which was recently observed in LBCO at 1/8th doping,\cite{NernstLBCO} along with measurements of an anomalous linear dichroism in YBCO and LBCO,\cite{PhysRevLett.112.147001} is consistent with such a mirror symmetry breaking phase, and is possible even without broken time-reversal symmetry.

Finally, let us discuss the relevance of these results to the doping dependence of charge order inferred from X-ray data in YBCO. Charge order has been seen to onset between 100K and 150K for hole concentrations in the range 7.8\% to 13.2$\%$.\cite{huecker2014competing} However, as previously mentioned, the X-ray peak magnitudes show a doping dependent anisotropy (which is minimal at one-eight doping).\cite{blanco2013momentum,blancoXray} Crucially, given that the orthorhombicity is small, strongly anisotropic peak magnitudes can \textit{only} be consistent with stripe like order. Thus, assuming universality of the charge ordering tendencies in YBCO, it is likely that the charge density waves are indeed stripe-like (and not checkerboard) \textit{at all dopings}. The role of interlayer couplings now becomes clear: subtle changes in the interlayer physics at different dopings or chain configurations can lead to a variety of stripe-like patterns.  It is therefore plausible that the asymmetry in X-ray peak heights in Ortho II samples is due to the presence of a parallel striped phase, while closer to one-eighth doping, small differences in interlayer couplings result in a crisscrossed state that gives roughly equal magnitude X-ray peaks in the $x$ and $y$ directions.

%----------------------------------------------------------------------------------------------------------------------------------------------------

\section{Conclusions} 
In summary, we have shown how unidirectional charge density waves in coupled bilayers can form either parallel or criss-phases depending on the form and strength of interlayer couplings and have suggested that these striped states provide a simple explanation for the observed doping dependence of charge order in YBCO. We have also argued that upon generalizing to multiple layers, this can lead to a variety of states which break mirror and $C_4$ rotation symmetries which may survive in the presence of quenched disorder. Indeed, while there is now a convincing case for short ranged incommensurate charge order in the pseudogap regime of the cuprates, there remains evidence of thermodynamic phase transitions to broken symmetry states at similar temperatures from probes such as the Kerr effect,\cite{KerrYBCO,He2011,KerrLBCO} whose relationship to charge order at present remains poorly understood.\cite{Hosur2}

\section{Acknowledgements}
We acknowledge helpful discussions with Philipp Dumitrescu, Marc-Henri Julien, Samuel Lederer, Brad Ramshaw, Elizabeth Schemm, Suchitra Sebastian, Yuxuan Wang, Jasper van Wezel, and especially Steven Kivelson. This work was supported by the U.S. DOE, Office of Basic Energy Sciences, under contract DEAC02-76SF00515 (AM, PH and SR), the Alfred P. Sloan Foundation (SR),  and the David and Lucile Packard Foundation (PH). 

\bibliography{tunneling}

\begin{thebibliography}{54}%
\makeatletter
\providecommand \@ifxundefined [1]{%
 \@ifx{#1\undefined}
}%
\providecommand \@ifnum [1]{%
 \ifnum #1\expandafter \@firstoftwo
 \else \expandafter \@secondoftwo
 \fi
}%
\providecommand \@ifx [1]{%
 \ifx #1\expandafter \@firstoftwo
 \else \expandafter \@secondoftwo
 \fi
}%
\providecommand \natexlab [1]{#1}%
\providecommand \enquote  [1]{``#1''}%
\providecommand \bibnamefont  [1]{#1}%
\providecommand \bibfnamefont [1]{#1}%
\providecommand \citenamefont [1]{#1}%
\providecommand \href@noop [0]{\@secondoftwo}%
\providecommand \href [0]{\begingroup \@sanitize@url \@href}%
\providecommand \@href[1]{\@@startlink{#1}\@@href}%
\providecommand \@@href[1]{\endgroup#1\@@endlink}%
\providecommand \@sanitize@url [0]{\catcode `\\12\catcode `\$12\catcode
  `\&12\catcode `\#12\catcode `\^12\catcode `\_12\catcode `\%12\relax}%
\providecommand \@@startlink[1]{}%
\providecommand \@@endlink[0]{}%
\providecommand \url  [0]{\begingroup\@sanitize@url \@url }%
\providecommand \@url [1]{\endgroup\@href {#1}{\urlprefix }}%
\providecommand \urlprefix  [0]{URL }%
\providecommand \Eprint [0]{\href }%
\providecommand \doibase [0]{http://dx.doi.org/}%
\providecommand \selectlanguage [0]{\@gobble}%
\providecommand \bibinfo  [0]{\@secondoftwo}%
\providecommand \bibfield  [0]{\@secondoftwo}%
\providecommand \translation [1]{[#1]}%
\providecommand \BibitemOpen [0]{}%
\providecommand \bibitemStop [0]{}%
\providecommand \bibitemNoStop [0]{.\EOS\space}%
\providecommand \EOS [0]{\spacefactor3000\relax}%
\providecommand \BibitemShut  [1]{\csname bibitem#1\endcsname}%
\let\auto@bib@innerbib\@empty
%</preamble>
\bibitem [{\citenamefont {{Chang}}\ \emph {et~al.}(2010)\citenamefont
  {{Chang}}, \citenamefont {{Daou}}, \citenamefont {{Proust}}, \citenamefont
  {{Leboeuf}}, \citenamefont {{Doiron-Leyraud}}, \citenamefont
  {{Lalibert{\'e}}}, \citenamefont {{Pingault}}, \citenamefont {{Ramshaw}},
  \citenamefont {{Liang}}, \citenamefont {{Bonn}}, \citenamefont {{Hardy}},
  \citenamefont {{Takagi}}, \citenamefont {{Antunes}}, \citenamefont
  {{Sheikin}}, \citenamefont {{Behnia}},\ and\ \citenamefont
  {{Taillefer}}}]{Chang2010}%
  \BibitemOpen
  \bibfield  {author} {\bibinfo {author} {\bibfnamefont {J.}~\bibnamefont
  {{Chang}}}, \bibinfo {author} {\bibfnamefont {R.}~\bibnamefont {{Daou}}},
  \bibinfo {author} {\bibfnamefont {C.}~\bibnamefont {{Proust}}}, \bibinfo
  {author} {\bibfnamefont {D.}~\bibnamefont {{Leboeuf}}}, \bibinfo {author}
  {\bibfnamefont {N.}~\bibnamefont {{Doiron-Leyraud}}}, \bibinfo {author}
  {\bibfnamefont {F.}~\bibnamefont {{Lalibert{\'e}}}}, \bibinfo {author}
  {\bibfnamefont {B.}~\bibnamefont {{Pingault}}}, \bibinfo {author}
  {\bibfnamefont {B.~J.}\ \bibnamefont {{Ramshaw}}}, \bibinfo {author}
  {\bibfnamefont {R.}~\bibnamefont {{Liang}}}, \bibinfo {author} {\bibfnamefont
  {D.~A.}\ \bibnamefont {{Bonn}}}, \bibinfo {author} {\bibfnamefont {W.~N.}\
  \bibnamefont {{Hardy}}}, \bibinfo {author} {\bibfnamefont {H.}~\bibnamefont
  {{Takagi}}}, \bibinfo {author} {\bibfnamefont {A.~B.}\ \bibnamefont
  {{Antunes}}}, \bibinfo {author} {\bibfnamefont {I.}~\bibnamefont
  {{Sheikin}}}, \bibinfo {author} {\bibfnamefont {K.}~\bibnamefont {{Behnia}}},
  \ and\ \bibinfo {author} {\bibfnamefont {L.}~\bibnamefont {{Taillefer}}},\
  }\href {\doibase 10.1103/PhysRevLett.104.057005} {\bibfield  {journal}
  {\bibinfo  {journal} {Physical Review Letters}\ }\textbf {\bibinfo {volume}
  {104}},\ \bibinfo {eid} {057005} (\bibinfo {year} {2010})},\ \Eprint
  {http://arxiv.org/abs/0907.5039} {arXiv:0907.5039 [cond-mat.supr-con]}
  \BibitemShut {NoStop}%
\bibitem [{\citenamefont {Ghiringhelli}\ \emph {et~al.}(2012)\citenamefont
  {Ghiringhelli}, \citenamefont {Le~Tacon}, \citenamefont {Minola},
  \citenamefont {Blanco-Canosa}, \citenamefont {Mazzoli}, \citenamefont
  {Brookes}, \citenamefont {De~Luca}, \citenamefont {Frano}, \citenamefont
  {Hawthorn}, \citenamefont {He} \emph {et~al.}}]{XrayYBCO}%
  \BibitemOpen
  \bibfield  {author} {\bibinfo {author} {\bibfnamefont {G.}~\bibnamefont
  {Ghiringhelli}}, \bibinfo {author} {\bibfnamefont {M.}~\bibnamefont
  {Le~Tacon}}, \bibinfo {author} {\bibfnamefont {M.}~\bibnamefont {Minola}},
  \bibinfo {author} {\bibfnamefont {S.}~\bibnamefont {Blanco-Canosa}}, \bibinfo
  {author} {\bibfnamefont {C.}~\bibnamefont {Mazzoli}}, \bibinfo {author}
  {\bibfnamefont {N.}~\bibnamefont {Brookes}}, \bibinfo {author} {\bibfnamefont
  {G.}~\bibnamefont {De~Luca}}, \bibinfo {author} {\bibfnamefont
  {A.}~\bibnamefont {Frano}}, \bibinfo {author} {\bibfnamefont
  {D.}~\bibnamefont {Hawthorn}}, \bibinfo {author} {\bibfnamefont
  {F.}~\bibnamefont {He}},  \emph {et~al.},\ }\href@noop {} {\bibfield
  {journal} {\bibinfo  {journal} {Science}\ }\textbf {\bibinfo {volume}
  {337}},\ \bibinfo {pages} {821} (\bibinfo {year} {2012})}\BibitemShut
  {NoStop}%
\bibitem [{\citenamefont {Chang}\ \emph {et~al.}(2012)\citenamefont {Chang},
  \citenamefont {Blackburn}, \citenamefont {Holmes}, \citenamefont
  {Christensen}, \citenamefont {Larsen}, \citenamefont {Mesot}, \citenamefont
  {Liang}, \citenamefont {Bonn}, \citenamefont {Hardy}, \citenamefont
  {Watenphul} \emph {et~al.}}]{chang2012direct}%
  \BibitemOpen
  \bibfield  {author} {\bibinfo {author} {\bibfnamefont {J.}~\bibnamefont
  {Chang}}, \bibinfo {author} {\bibfnamefont {E.}~\bibnamefont {Blackburn}},
  \bibinfo {author} {\bibfnamefont {A.}~\bibnamefont {Holmes}}, \bibinfo
  {author} {\bibfnamefont {N.}~\bibnamefont {Christensen}}, \bibinfo {author}
  {\bibfnamefont {J.}~\bibnamefont {Larsen}}, \bibinfo {author} {\bibfnamefont
  {J.}~\bibnamefont {Mesot}}, \bibinfo {author} {\bibfnamefont
  {R.}~\bibnamefont {Liang}}, \bibinfo {author} {\bibfnamefont
  {D.}~\bibnamefont {Bonn}}, \bibinfo {author} {\bibfnamefont {W.}~\bibnamefont
  {Hardy}}, \bibinfo {author} {\bibfnamefont {A.}~\bibnamefont {Watenphul}},
  \emph {et~al.},\ }\href@noop {} {\bibfield  {journal} {\bibinfo  {journal}
  {Nature Physics}\ }\textbf {\bibinfo {volume} {8}},\ \bibinfo {pages} {871}
  (\bibinfo {year} {2012})}\BibitemShut {NoStop}%
\bibitem [{\citenamefont {Achkar}\ \emph {et~al.}(2012)\citenamefont {Achkar},
  \citenamefont {Sutarto}, \citenamefont {Mao}, \citenamefont {He},
  \citenamefont {Frano}, \citenamefont {Blanco-Canosa}, \citenamefont
  {Le~Tacon}, \citenamefont {Ghiringhelli}, \citenamefont {Braicovich},
  \citenamefont {Minola}, \citenamefont {Moretti~Sala}, \citenamefont
  {Mazzoli}, \citenamefont {Liang}, \citenamefont {Bonn}, \citenamefont
  {Hardy}, \citenamefont {Keimer}, \citenamefont {Sawatzky},\ and\
  \citenamefont {Hawthorn}}]{RexsYBCO}%
  \BibitemOpen
  \bibfield  {author} {\bibinfo {author} {\bibfnamefont {A.~J.}\ \bibnamefont
  {Achkar}}, \bibinfo {author} {\bibfnamefont {R.}~\bibnamefont {Sutarto}},
  \bibinfo {author} {\bibfnamefont {X.}~\bibnamefont {Mao}}, \bibinfo {author}
  {\bibfnamefont {F.}~\bibnamefont {He}}, \bibinfo {author} {\bibfnamefont
  {A.}~\bibnamefont {Frano}}, \bibinfo {author} {\bibfnamefont
  {S.}~\bibnamefont {Blanco-Canosa}}, \bibinfo {author} {\bibfnamefont
  {M.}~\bibnamefont {Le~Tacon}}, \bibinfo {author} {\bibfnamefont
  {G.}~\bibnamefont {Ghiringhelli}}, \bibinfo {author} {\bibfnamefont
  {L.}~\bibnamefont {Braicovich}}, \bibinfo {author} {\bibfnamefont
  {M.}~\bibnamefont {Minola}}, \bibinfo {author} {\bibfnamefont
  {M.}~\bibnamefont {Moretti~Sala}}, \bibinfo {author} {\bibfnamefont
  {C.}~\bibnamefont {Mazzoli}}, \bibinfo {author} {\bibfnamefont
  {R.}~\bibnamefont {Liang}}, \bibinfo {author} {\bibfnamefont {D.~A.}\
  \bibnamefont {Bonn}}, \bibinfo {author} {\bibfnamefont {W.~N.}\ \bibnamefont
  {Hardy}}, \bibinfo {author} {\bibfnamefont {B.}~\bibnamefont {Keimer}},
  \bibinfo {author} {\bibfnamefont {G.~A.}\ \bibnamefont {Sawatzky}}, \ and\
  \bibinfo {author} {\bibfnamefont {D.~G.}\ \bibnamefont {Hawthorn}},\ }\href
  {\doibase 10.1103/PhysRevLett.109.167001} {\bibfield  {journal} {\bibinfo
  {journal} {Phys. Rev. Lett.}\ }\textbf {\bibinfo {volume} {109}},\ \bibinfo
  {pages} {167001} (\bibinfo {year} {2012})}\BibitemShut {NoStop}%
\bibitem [{\citenamefont {Wu}\ \emph {et~al.}(2014)\citenamefont {Wu},
  \citenamefont {Mayaffre}, \citenamefont {Kr{\"a}mer}, \citenamefont
  {Horvati{\'c}}, \citenamefont {Berthier}, \citenamefont {Hardy},
  \citenamefont {Liang}, \citenamefont {Bonn},\ and\ \citenamefont
  {Julien}}]{wu2014short}%
  \BibitemOpen
  \bibfield  {author} {\bibinfo {author} {\bibfnamefont {T.}~\bibnamefont
  {Wu}}, \bibinfo {author} {\bibfnamefont {H.}~\bibnamefont {Mayaffre}},
  \bibinfo {author} {\bibfnamefont {S.}~\bibnamefont {Kr{\"a}mer}}, \bibinfo
  {author} {\bibfnamefont {M.}~\bibnamefont {Horvati{\'c}}}, \bibinfo {author}
  {\bibfnamefont {C.}~\bibnamefont {Berthier}}, \bibinfo {author}
  {\bibfnamefont {W.}~\bibnamefont {Hardy}}, \bibinfo {author} {\bibfnamefont
  {R.}~\bibnamefont {Liang}}, \bibinfo {author} {\bibfnamefont
  {D.}~\bibnamefont {Bonn}}, \ and\ \bibinfo {author} {\bibfnamefont {M.-H.}\
  \bibnamefont {Julien}},\ }\href@noop {} {\bibfield  {journal} {\bibinfo
  {journal} {arXiv preprint arXiv:1404.1617}\ } (\bibinfo {year}
  {2014})}\BibitemShut {NoStop}%
\bibitem [{\citenamefont {Wu}\ \emph {et~al.}(2011)\citenamefont {Wu},
  \citenamefont {Mayaffre}, \citenamefont {Kramer}, \citenamefont {Horvatic},
  \citenamefont {Berthier}, \citenamefont {Hardy}, \citenamefont {Liang},
  \citenamefont {Bonn},\ and\ \citenamefont {Julien}}]{NMRYBCO}%
  \BibitemOpen
  \bibfield  {author} {\bibinfo {author} {\bibfnamefont {T.}~\bibnamefont
  {Wu}}, \bibinfo {author} {\bibfnamefont {H.}~\bibnamefont {Mayaffre}},
  \bibinfo {author} {\bibfnamefont {S.}~\bibnamefont {Kramer}}, \bibinfo
  {author} {\bibfnamefont {M.}~\bibnamefont {Horvatic}}, \bibinfo {author}
  {\bibfnamefont {C.}~\bibnamefont {Berthier}}, \bibinfo {author}
  {\bibfnamefont {W.}~\bibnamefont {Hardy}}, \bibinfo {author} {\bibfnamefont
  {R.}~\bibnamefont {Liang}}, \bibinfo {author} {\bibfnamefont
  {D.}~\bibnamefont {Bonn}}, \ and\ \bibinfo {author} {\bibfnamefont {M.-H.}\
  \bibnamefont {Julien}},\ }\href@noop {} {\bibfield  {journal} {\bibinfo
  {journal} {Nature}\ }\textbf {\bibinfo {volume} {477}},\ \bibinfo {pages}
  {191} (\bibinfo {year} {2011})}\BibitemShut {NoStop}%
\bibitem [{\citenamefont {Doiron-Leyraud}\ \emph {et~al.}(2007)\citenamefont
  {Doiron-Leyraud}, \citenamefont {Proust}, \citenamefont {LeBoeuf},
  \citenamefont {Levallois}, \citenamefont {Bonnemaison}, \citenamefont
  {Liang}, \citenamefont {Bonn}, \citenamefont {Hardy},\ and\ \citenamefont
  {Taillefer}}]{Proust2007}%
  \BibitemOpen
  \bibfield  {author} {\bibinfo {author} {\bibfnamefont {N.}~\bibnamefont
  {Doiron-Leyraud}}, \bibinfo {author} {\bibfnamefont {C.}~\bibnamefont
  {Proust}}, \bibinfo {author} {\bibfnamefont {D.}~\bibnamefont {LeBoeuf}},
  \bibinfo {author} {\bibfnamefont {J.}~\bibnamefont {Levallois}}, \bibinfo
  {author} {\bibfnamefont {J.}~\bibnamefont {Bonnemaison}}, \bibinfo {author}
  {\bibfnamefont {R.}~\bibnamefont {Liang}}, \bibinfo {author} {\bibfnamefont
  {D.}~\bibnamefont {Bonn}}, \bibinfo {author} {\bibfnamefont {W.}~\bibnamefont
  {Hardy}}, \ and\ \bibinfo {author} {\bibfnamefont {L.}~\bibnamefont
  {Taillefer}},\ }\href@noop {} {\bibfield  {journal} {\bibinfo  {journal}
  {Nature}\ }\textbf {\bibinfo {volume} {447}},\ \bibinfo {pages} {565}
  (\bibinfo {year} {2007})}\BibitemShut {NoStop}%
\bibitem [{\citenamefont {LeBoeuf}\ \emph {et~al.}(2007)\citenamefont
  {LeBoeuf}, \citenamefont {Doiron-Leyraud}, \citenamefont {Levallois},
  \citenamefont {Daou}, \citenamefont {Bonnemaison}, \citenamefont {Hussey},
  \citenamefont {Balicas}, \citenamefont {Ramshaw}, \citenamefont {Liang},
  \citenamefont {Bonn} \emph {et~al.}}]{Proust2007a}%
  \BibitemOpen
  \bibfield  {author} {\bibinfo {author} {\bibfnamefont {D.}~\bibnamefont
  {LeBoeuf}}, \bibinfo {author} {\bibfnamefont {N.}~\bibnamefont
  {Doiron-Leyraud}}, \bibinfo {author} {\bibfnamefont {J.}~\bibnamefont
  {Levallois}}, \bibinfo {author} {\bibfnamefont {R.}~\bibnamefont {Daou}},
  \bibinfo {author} {\bibfnamefont {J.}~\bibnamefont {Bonnemaison}}, \bibinfo
  {author} {\bibfnamefont {N.}~\bibnamefont {Hussey}}, \bibinfo {author}
  {\bibfnamefont {L.}~\bibnamefont {Balicas}}, \bibinfo {author} {\bibfnamefont
  {B.}~\bibnamefont {Ramshaw}}, \bibinfo {author} {\bibfnamefont
  {R.}~\bibnamefont {Liang}}, \bibinfo {author} {\bibfnamefont
  {D.}~\bibnamefont {Bonn}},  \emph {et~al.},\ }\href@noop {} {\bibfield
  {journal} {\bibinfo  {journal} {Nature}\ }\textbf {\bibinfo {volume} {450}},\
  \bibinfo {pages} {533} (\bibinfo {year} {2007})}\BibitemShut {NoStop}%
\bibitem [{\citenamefont {Sebastian}\ \emph {et~al.}(2009)\citenamefont
  {Sebastian}, \citenamefont {Harrison}, \citenamefont {Mielke}, \citenamefont
  {Liang}, \citenamefont {Bonn}, \citenamefont {Hardy},\ and\ \citenamefont
  {Lonzarich}}]{Sebastian2009}%
  \BibitemOpen
  \bibfield  {author} {\bibinfo {author} {\bibfnamefont {S.~E.}\ \bibnamefont
  {Sebastian}}, \bibinfo {author} {\bibfnamefont {N.}~\bibnamefont {Harrison}},
  \bibinfo {author} {\bibfnamefont {C.~H.}\ \bibnamefont {Mielke}}, \bibinfo
  {author} {\bibfnamefont {R.}~\bibnamefont {Liang}}, \bibinfo {author}
  {\bibfnamefont {D.~A.}\ \bibnamefont {Bonn}}, \bibinfo {author}
  {\bibfnamefont {W.~N.}\ \bibnamefont {Hardy}}, \ and\ \bibinfo {author}
  {\bibfnamefont {G.~G.}\ \bibnamefont {Lonzarich}},\ }\href {\doibase
  {10.1103/PhysRevLett.103.256405}} {\bibfield  {journal} {\bibinfo  {journal}
  {{Physical Review Letters}}\ }\textbf {\bibinfo {volume} {{103}}} (\bibinfo
  {year} {{2009}}),\ {10.1103/PhysRevLett.103.256405}}\BibitemShut {NoStop}%
\bibitem [{\citenamefont {Sebastian}\ \emph {et~al.}(2010)\citenamefont
  {Sebastian}, \citenamefont {Harrison}, \citenamefont {Altarawneh},
  \citenamefont {Liang}, \citenamefont {Bonn}, \citenamefont {Hardy},\ and\
  \citenamefont {Lonzarich}}]{Sebastian2010}%
  \BibitemOpen
  \bibfield  {author} {\bibinfo {author} {\bibfnamefont {S.~E.}\ \bibnamefont
  {Sebastian}}, \bibinfo {author} {\bibfnamefont {N.}~\bibnamefont {Harrison}},
  \bibinfo {author} {\bibfnamefont {M.~M.}\ \bibnamefont {Altarawneh}},
  \bibinfo {author} {\bibfnamefont {R.}~\bibnamefont {Liang}}, \bibinfo
  {author} {\bibfnamefont {D.~A.}\ \bibnamefont {Bonn}}, \bibinfo {author}
  {\bibfnamefont {W.~N.}\ \bibnamefont {Hardy}}, \ and\ \bibinfo {author}
  {\bibfnamefont {G.~G.}\ \bibnamefont {Lonzarich}},\ }\href {\doibase
  {10.1103/PhysRevB.81.140505}} {\bibfield  {journal} {\bibinfo  {journal}
  {{Physical Review B}}\ }\textbf {\bibinfo {volume} {{81}}} (\bibinfo {year}
  {{2010}}),\ {10.1103/PhysRevB.81.140505}}\BibitemShut {NoStop}%
\bibitem [{\citenamefont {Sebastian}\ \emph {et~al.}(2012)\citenamefont
  {Sebastian}, \citenamefont {Harrison},\ and\ \citenamefont
  {Lonzarich}}]{sebastian2012towards}%
  \BibitemOpen
  \bibfield  {author} {\bibinfo {author} {\bibfnamefont {S.~E.}\ \bibnamefont
  {Sebastian}}, \bibinfo {author} {\bibfnamefont {N.}~\bibnamefont {Harrison}},
  \ and\ \bibinfo {author} {\bibfnamefont {G.~G.}\ \bibnamefont {Lonzarich}},\
  }\href@noop {} {\bibfield  {journal} {\bibinfo  {journal} {Reports on
  Progress in Physics}\ }\textbf {\bibinfo {volume} {75}},\ \bibinfo {pages}
  {102501} (\bibinfo {year} {2012})}\BibitemShut {NoStop}%
\bibitem [{\citenamefont {H\"ucker}\ \emph {et~al.}(2014)\citenamefont
  {H\"ucker}, \citenamefont {Christensen}, \citenamefont {Holmes},
  \citenamefont {Blackburn}, \citenamefont {Forgan}, \citenamefont {Liang},
  \citenamefont {Bonn}, \citenamefont {Hardy}, \citenamefont {Gutowski},
  \citenamefont {Zimmermann}, \citenamefont {Hayden},\ and\ \citenamefont
  {Chang}}]{huecker2014competing}%
  \BibitemOpen
  \bibfield  {author} {\bibinfo {author} {\bibfnamefont {M.}~\bibnamefont
  {H\"ucker}}, \bibinfo {author} {\bibfnamefont {N.~B.}\ \bibnamefont
  {Christensen}}, \bibinfo {author} {\bibfnamefont {A.~T.}\ \bibnamefont
  {Holmes}}, \bibinfo {author} {\bibfnamefont {E.}~\bibnamefont {Blackburn}},
  \bibinfo {author} {\bibfnamefont {E.~M.}\ \bibnamefont {Forgan}}, \bibinfo
  {author} {\bibfnamefont {R.}~\bibnamefont {Liang}}, \bibinfo {author}
  {\bibfnamefont {D.~A.}\ \bibnamefont {Bonn}}, \bibinfo {author}
  {\bibfnamefont {W.~N.}\ \bibnamefont {Hardy}}, \bibinfo {author}
  {\bibfnamefont {O.}~\bibnamefont {Gutowski}}, \bibinfo {author}
  {\bibfnamefont {M.~v.}\ \bibnamefont {Zimmermann}}, \bibinfo {author}
  {\bibfnamefont {S.~M.}\ \bibnamefont {Hayden}}, \ and\ \bibinfo {author}
  {\bibfnamefont {J.}~\bibnamefont {Chang}},\ }\href {\doibase
  10.1103/PhysRevB.90.054514} {\bibfield  {journal} {\bibinfo  {journal} {Phys.
  Rev. B}\ }\textbf {\bibinfo {volume} {90}},\ \bibinfo {pages} {054514}
  (\bibinfo {year} {2014})}\BibitemShut {NoStop}%
\bibitem [{\citenamefont {Blanco-Canosa}\ \emph {et~al.}(2014)\citenamefont
  {Blanco-Canosa}, \citenamefont {Frano}, \citenamefont {Schierle},
  \citenamefont {Porras}, \citenamefont {Loew}, \citenamefont {Minola},
  \citenamefont {Bluschke}, \citenamefont {Weschke}, \citenamefont {Keimer},\
  and\ \citenamefont {Le~Tacon}}]{blancoXray}%
  \BibitemOpen
  \bibfield  {author} {\bibinfo {author} {\bibfnamefont {S.}~\bibnamefont
  {Blanco-Canosa}}, \bibinfo {author} {\bibfnamefont {A.}~\bibnamefont
  {Frano}}, \bibinfo {author} {\bibfnamefont {E.}~\bibnamefont {Schierle}},
  \bibinfo {author} {\bibfnamefont {J.}~\bibnamefont {Porras}}, \bibinfo
  {author} {\bibfnamefont {T.}~\bibnamefont {Loew}}, \bibinfo {author}
  {\bibfnamefont {M.}~\bibnamefont {Minola}}, \bibinfo {author} {\bibfnamefont
  {M.}~\bibnamefont {Bluschke}}, \bibinfo {author} {\bibfnamefont
  {E.}~\bibnamefont {Weschke}}, \bibinfo {author} {\bibfnamefont
  {B.}~\bibnamefont {Keimer}}, \ and\ \bibinfo {author} {\bibfnamefont
  {M.}~\bibnamefont {Le~Tacon}},\ }\href {\doibase 10.1103/PhysRevB.90.054513}
  {\bibfield  {journal} {\bibinfo  {journal} {Phys. Rev. B}\ }\textbf {\bibinfo
  {volume} {90}},\ \bibinfo {pages} {054513} (\bibinfo {year}
  {2014})}\BibitemShut {NoStop}%
\bibitem [{\citenamefont {Blanco-Canosa}\ \emph {et~al.}(2013)\citenamefont
  {Blanco-Canosa}, \citenamefont {Frano}, \citenamefont {Loew}, \citenamefont
  {Lu}, \citenamefont {Porras}, \citenamefont {Ghiringhelli}, \citenamefont
  {Minola}, \citenamefont {Mazzoli}, \citenamefont {Braicovich}, \citenamefont
  {Schierle} \emph {et~al.}}]{blanco2013momentum}%
  \BibitemOpen
  \bibfield  {author} {\bibinfo {author} {\bibfnamefont {S.}~\bibnamefont
  {Blanco-Canosa}}, \bibinfo {author} {\bibfnamefont {A.}~\bibnamefont
  {Frano}}, \bibinfo {author} {\bibfnamefont {T.}~\bibnamefont {Loew}},
  \bibinfo {author} {\bibfnamefont {Y.}~\bibnamefont {Lu}}, \bibinfo {author}
  {\bibfnamefont {J.}~\bibnamefont {Porras}}, \bibinfo {author} {\bibfnamefont
  {G.}~\bibnamefont {Ghiringhelli}}, \bibinfo {author} {\bibfnamefont
  {M.}~\bibnamefont {Minola}}, \bibinfo {author} {\bibfnamefont
  {C.}~\bibnamefont {Mazzoli}}, \bibinfo {author} {\bibfnamefont
  {L.}~\bibnamefont {Braicovich}}, \bibinfo {author} {\bibfnamefont
  {E.}~\bibnamefont {Schierle}},  \emph {et~al.},\ }\href@noop {} {\bibfield
  {journal} {\bibinfo  {journal} {Physical Review Letters}\ }\textbf {\bibinfo
  {volume} {110}},\ \bibinfo {pages} {187001} (\bibinfo {year}
  {2013})}\BibitemShut {NoStop}%
\bibitem [{\citenamefont {Fujita}\ \emph {et~al.}(2004)\citenamefont {Fujita},
  \citenamefont {Goka}, \citenamefont {Yamada}, \citenamefont {Tranquada},\
  and\ \citenamefont {Regnault}}]{Tranquada2004}%
  \BibitemOpen
  \bibfield  {author} {\bibinfo {author} {\bibfnamefont {M.}~\bibnamefont
  {Fujita}}, \bibinfo {author} {\bibfnamefont {H.}~\bibnamefont {Goka}},
  \bibinfo {author} {\bibfnamefont {K.}~\bibnamefont {Yamada}}, \bibinfo
  {author} {\bibfnamefont {J.~M.}\ \bibnamefont {Tranquada}}, \ and\ \bibinfo
  {author} {\bibfnamefont {L.~P.}\ \bibnamefont {Regnault}},\ }\href {\doibase
  10.1103/PhysRevB.70.104517} {\bibfield  {journal} {\bibinfo  {journal} {Phys.
  Rev. B}\ }\textbf {\bibinfo {volume} {70}},\ \bibinfo {pages} {104517}
  (\bibinfo {year} {2004})}\BibitemShut {NoStop}%
\bibitem [{\citenamefont {Suzuki}\ \emph {et~al.}(1998)\citenamefont {Suzuki},
  \citenamefont {Goto}, \citenamefont {Chiba}, \citenamefont {Shinoda},
  \citenamefont {Fukase}, \citenamefont {Kimura}, \citenamefont {Yamada},
  \citenamefont {Ohashi},\ and\ \citenamefont
  {Yamaguchi}}]{suzuki1998observation}%
  \BibitemOpen
  \bibfield  {author} {\bibinfo {author} {\bibfnamefont {T.}~\bibnamefont
  {Suzuki}}, \bibinfo {author} {\bibfnamefont {T.}~\bibnamefont {Goto}},
  \bibinfo {author} {\bibfnamefont {K.}~\bibnamefont {Chiba}}, \bibinfo
  {author} {\bibfnamefont {T.}~\bibnamefont {Shinoda}}, \bibinfo {author}
  {\bibfnamefont {T.}~\bibnamefont {Fukase}}, \bibinfo {author} {\bibfnamefont
  {H.}~\bibnamefont {Kimura}}, \bibinfo {author} {\bibfnamefont
  {K.}~\bibnamefont {Yamada}}, \bibinfo {author} {\bibfnamefont
  {M.}~\bibnamefont {Ohashi}}, \ and\ \bibinfo {author} {\bibfnamefont
  {Y.}~\bibnamefont {Yamaguchi}},\ }\href@noop {} {\bibfield  {journal}
  {\bibinfo  {journal} {Physical Review B}\ }\textbf {\bibinfo {volume} {57}},\
  \bibinfo {pages} {R3229} (\bibinfo {year} {1998})}\BibitemShut {NoStop}%
\bibitem [{\citenamefont {Howald}\ \emph {et~al.}(2003)\citenamefont {Howald},
  \citenamefont {Eisaki}, \citenamefont {Kaneko}, \citenamefont {Greven},\ and\
  \citenamefont {Kapitulnik}}]{howald2003periodic}%
  \BibitemOpen
  \bibfield  {author} {\bibinfo {author} {\bibfnamefont {C.}~\bibnamefont
  {Howald}}, \bibinfo {author} {\bibfnamefont {H.}~\bibnamefont {Eisaki}},
  \bibinfo {author} {\bibfnamefont {N.}~\bibnamefont {Kaneko}}, \bibinfo
  {author} {\bibfnamefont {M.}~\bibnamefont {Greven}}, \ and\ \bibinfo {author}
  {\bibfnamefont {A.}~\bibnamefont {Kapitulnik}},\ }\href@noop {} {\bibfield
  {journal} {\bibinfo  {journal} {Physical Review B}\ }\textbf {\bibinfo
  {volume} {67}},\ \bibinfo {pages} {014533} (\bibinfo {year}
  {2003})}\BibitemShut {NoStop}%
\bibitem [{Note1()}]{Note1}%
  \BibitemOpen
  \bibinfo {note} {There are many subtle issues in distinguishing striped from
  checkerboard order from X-ray data when the order is not long ranged. See
  Sec.V of the Supplemental Material of \protect \citep {nie2013quenched} also
  Ref.~\protect \rev@citealpnum {robertson2006distinguishing}}\BibitemShut
  {NoStop}%
\bibitem [{\citenamefont {Fink}\ \emph {et~al.}(2009)\citenamefont {Fink},
  \citenamefont {Schierle}, \citenamefont {Weschke}, \citenamefont {Geck},
  \citenamefont {Hawthorn}, \citenamefont {Soltwisch}, \citenamefont {Wadati},
  \citenamefont {Wu}, \citenamefont {D\"urr}, \citenamefont {Wizent},
  \citenamefont {B\"uchner},\ and\ \citenamefont {Sawatzky}}]{Tranquada2009}%
  \BibitemOpen
  \bibfield  {author} {\bibinfo {author} {\bibfnamefont {J.}~\bibnamefont
  {Fink}}, \bibinfo {author} {\bibfnamefont {E.}~\bibnamefont {Schierle}},
  \bibinfo {author} {\bibfnamefont {E.}~\bibnamefont {Weschke}}, \bibinfo
  {author} {\bibfnamefont {J.}~\bibnamefont {Geck}}, \bibinfo {author}
  {\bibfnamefont {D.}~\bibnamefont {Hawthorn}}, \bibinfo {author}
  {\bibfnamefont {V.}~\bibnamefont {Soltwisch}}, \bibinfo {author}
  {\bibfnamefont {H.}~\bibnamefont {Wadati}}, \bibinfo {author} {\bibfnamefont
  {H.-H.}\ \bibnamefont {Wu}}, \bibinfo {author} {\bibfnamefont {H.~A.}\
  \bibnamefont {D\"urr}}, \bibinfo {author} {\bibfnamefont {N.}~\bibnamefont
  {Wizent}}, \bibinfo {author} {\bibfnamefont {B.}~\bibnamefont {B\"uchner}}, \
  and\ \bibinfo {author} {\bibfnamefont {G.~A.}\ \bibnamefont {Sawatzky}},\
  }\href {\doibase 10.1103/PhysRevB.79.100502} {\bibfield  {journal} {\bibinfo
  {journal} {Phys. Rev. B}\ }\textbf {\bibinfo {volume} {79}},\ \bibinfo
  {pages} {100502} (\bibinfo {year} {2009})}\BibitemShut {NoStop}%
\bibitem [{\citenamefont {Chakravarty}\ \emph {et~al.}(1993)\citenamefont
  {Chakravarty}, \citenamefont {Sudb{\o}}, \citenamefont {Anderson},\ and\
  \citenamefont {Strong}}]{chakravarty1993interlayer}%
  \BibitemOpen
  \bibfield  {author} {\bibinfo {author} {\bibfnamefont {S.}~\bibnamefont
  {Chakravarty}}, \bibinfo {author} {\bibfnamefont {A.}~\bibnamefont
  {Sudb{\o}}}, \bibinfo {author} {\bibfnamefont {P.~W.}\ \bibnamefont
  {Anderson}}, \ and\ \bibinfo {author} {\bibfnamefont {S.}~\bibnamefont
  {Strong}},\ }\href@noop {} {\bibfield  {journal} {\bibinfo  {journal}
  {Science}\ }\textbf {\bibinfo {volume} {261}},\ \bibinfo {pages} {337}
  (\bibinfo {year} {1993})}\BibitemShut {NoStop}%
\bibitem [{\citenamefont {Pavarini}\ \emph {et~al.}(2001)\citenamefont
  {Pavarini}, \citenamefont {Dasgupta}, \citenamefont {Saha-Dasgupta},
  \citenamefont {Jepsen},\ and\ \citenamefont {Andersen}}]{OkAndersen}%
  \BibitemOpen
  \bibfield  {author} {\bibinfo {author} {\bibfnamefont {E.}~\bibnamefont
  {Pavarini}}, \bibinfo {author} {\bibfnamefont {I.}~\bibnamefont {Dasgupta}},
  \bibinfo {author} {\bibfnamefont {T.}~\bibnamefont {Saha-Dasgupta}}, \bibinfo
  {author} {\bibfnamefont {O.}~\bibnamefont {Jepsen}}, \ and\ \bibinfo {author}
  {\bibfnamefont {O.}~\bibnamefont {Andersen}},\ }\href@noop {} {\bibfield
  {journal} {\bibinfo  {journal} {Physical Review Letters}\ }\textbf {\bibinfo
  {volume} {87}},\ \bibinfo {pages} {047003} (\bibinfo {year}
  {2001})}\BibitemShut {NoStop}%
\bibitem [{\citenamefont {Zaanen}\ and\ \citenamefont
  {Gunnarsson}(1989)}]{zaanen1989charged}%
  \BibitemOpen
  \bibfield  {author} {\bibinfo {author} {\bibfnamefont {J.}~\bibnamefont
  {Zaanen}}\ and\ \bibinfo {author} {\bibfnamefont {O.}~\bibnamefont
  {Gunnarsson}},\ }\href@noop {} {\bibfield  {journal} {\bibinfo  {journal}
  {Physical Review B}\ }\textbf {\bibinfo {volume} {40}},\ \bibinfo {pages}
  {7391} (\bibinfo {year} {1989})}\BibitemShut {NoStop}%
\bibitem [{\citenamefont {Machida}(1989)}]{machida1989magnetism}%
  \BibitemOpen
  \bibfield  {author} {\bibinfo {author} {\bibfnamefont {K.}~\bibnamefont
  {Machida}},\ }\href@noop {} {\bibfield  {journal} {\bibinfo  {journal}
  {Physica C: Superconductivity}\ }\textbf {\bibinfo {volume} {158}},\ \bibinfo
  {pages} {192} (\bibinfo {year} {1989})}\BibitemShut {NoStop}%
\bibitem [{\citenamefont {Schulz}(1990)}]{schulz1990incommensurate}%
  \BibitemOpen
  \bibfield  {author} {\bibinfo {author} {\bibfnamefont {H.~J.}\ \bibnamefont
  {Schulz}},\ }\href {\doibase 10.1103/PhysRevLett.64.1445} {\bibfield
  {journal} {\bibinfo  {journal} {Phys. Rev. Lett.}\ }\textbf {\bibinfo
  {volume} {64}},\ \bibinfo {pages} {1445} (\bibinfo {year}
  {1990})}\BibitemShut {NoStop}%
\bibitem [{\citenamefont {Emery}\ and\ \citenamefont
  {Kivelson}(1993)}]{emery1993frustrated}%
  \BibitemOpen
  \bibfield  {author} {\bibinfo {author} {\bibfnamefont {V.}~\bibnamefont
  {Emery}}\ and\ \bibinfo {author} {\bibfnamefont {S.}~\bibnamefont
  {Kivelson}},\ }\href@noop {} {\bibfield  {journal} {\bibinfo  {journal}
  {Physica C: Superconductivity}\ }\textbf {\bibinfo {volume} {209}},\ \bibinfo
  {pages} {597} (\bibinfo {year} {1993})}\BibitemShut {NoStop}%
\bibitem [{\citenamefont {Emery}\ \emph {et~al.}(1999)\citenamefont {Emery},
  \citenamefont {Kivelson},\ and\ \citenamefont {Tranquada}}]{emery1999stripe}%
  \BibitemOpen
  \bibfield  {author} {\bibinfo {author} {\bibfnamefont {V.}~\bibnamefont
  {Emery}}, \bibinfo {author} {\bibfnamefont {S.}~\bibnamefont {Kivelson}}, \
  and\ \bibinfo {author} {\bibfnamefont {J.}~\bibnamefont {Tranquada}},\
  }\href@noop {} {\bibfield  {journal} {\bibinfo  {journal} {Proceedings of the
  National Academy of Sciences}\ }\textbf {\bibinfo {volume} {96}},\ \bibinfo
  {pages} {8814} (\bibinfo {year} {1999})}\BibitemShut {NoStop}%
\bibitem [{\citenamefont {White}\ and\ \citenamefont
  {Scalapino}(1998)}]{white1998density}%
  \BibitemOpen
  \bibfield  {author} {\bibinfo {author} {\bibfnamefont {S.~R.}\ \bibnamefont
  {White}}\ and\ \bibinfo {author} {\bibfnamefont {D.}~\bibnamefont
  {Scalapino}},\ }\href@noop {} {\bibfield  {journal} {\bibinfo  {journal}
  {Physical Review Letters}\ }\textbf {\bibinfo {volume} {80}},\ \bibinfo
  {pages} {1272} (\bibinfo {year} {1998})}\BibitemShut {NoStop}%
\bibitem [{\citenamefont {Chakravarty}\ \emph {et~al.}(2001)\citenamefont
  {Chakravarty}, \citenamefont {Laughlin}, \citenamefont {Morr},\ and\
  \citenamefont {Nayak}}]{chakravarty2001hidden}%
  \BibitemOpen
  \bibfield  {author} {\bibinfo {author} {\bibfnamefont {S.}~\bibnamefont
  {Chakravarty}}, \bibinfo {author} {\bibfnamefont {R.}~\bibnamefont
  {Laughlin}}, \bibinfo {author} {\bibfnamefont {D.~K.}\ \bibnamefont {Morr}},
  \ and\ \bibinfo {author} {\bibfnamefont {C.}~\bibnamefont {Nayak}},\
  }\href@noop {} {\bibfield  {journal} {\bibinfo  {journal} {Physical Review
  B}\ }\textbf {\bibinfo {volume} {63}},\ \bibinfo {pages} {094503} (\bibinfo
  {year} {2001})}\BibitemShut {NoStop}%
\bibitem [{\citenamefont {Sachdev}\ and\ \citenamefont
  {La~Placa}(2013)}]{sachdev2013bond}%
  \BibitemOpen
  \bibfield  {author} {\bibinfo {author} {\bibfnamefont {S.}~\bibnamefont
  {Sachdev}}\ and\ \bibinfo {author} {\bibfnamefont {R.}~\bibnamefont
  {La~Placa}},\ }\href@noop {} {\bibfield  {journal} {\bibinfo  {journal}
  {Physical Review Letters}\ }\textbf {\bibinfo {volume} {111}},\ \bibinfo
  {pages} {027202} (\bibinfo {year} {2013})}\BibitemShut {NoStop}%
\bibitem [{\citenamefont {Wang}\ and\ \citenamefont
  {Chubukov}(2014)}]{wang2014charge}%
  \BibitemOpen
  \bibfield  {author} {\bibinfo {author} {\bibfnamefont {Y.}~\bibnamefont
  {Wang}}\ and\ \bibinfo {author} {\bibfnamefont {A.~V.}\ \bibnamefont
  {Chubukov}},\ }\href@noop {} {\bibfield  {journal} {\bibinfo  {journal}
  {arXiv preprint arXiv:1401.0712}\ } (\bibinfo {year} {2014})}\BibitemShut
  {NoStop}%
\bibitem [{\citenamefont {Laughlin}(2014)}]{laughlin2014fermi}%
  \BibitemOpen
  \bibfield  {author} {\bibinfo {author} {\bibfnamefont {R.}~\bibnamefont
  {Laughlin}},\ }\href@noop {} {\bibfield  {journal} {\bibinfo  {journal}
  {Physical Review Letters}\ }\textbf {\bibinfo {volume} {112}},\ \bibinfo
  {pages} {017004} (\bibinfo {year} {2014})}\BibitemShut {NoStop}%
\bibitem [{Note2()}]{Note2}%
  \BibitemOpen
  \bibinfo {note} {The existence of various types of density wave order, both
  at zero and finite temperature in the single layer model has been previously
  studied in the presence of incommensuration\cite
  {DelMaestro1,DelMaestro2}}\BibitemShut {NoStop}%
\bibitem [{Note3()}]{Note3}%
  \BibitemOpen
  \bibinfo {note} {While a period 3 density wave is close to the experimentally
  observed ordering vector of $|\protect \ensuremath \protect \bm {Q}|\approx
  2\pi \cdot 0.31$, we have chosen low order commensurate density waves for
  technical convenience as this greatly simplifies the sums over the reduced
  Brillouin zone (e.g. a period 3 density wave in both $x$ and $y$ directions
  corresponds to a Brillouin zone (BZ) that is 1/9th the original BZ. This
  means that the two layer Hamiltonian is an $18\times 18$ matrix, which is
  relatively simple to numerically analyze).}\BibitemShut {Stop}%
\bibitem [{\citenamefont {Eun}\ \emph {et~al.}(2012)\citenamefont {Eun},
  \citenamefont {Wang},\ and\ \citenamefont {Chakravarty}}]{eun2012quantum}%
  \BibitemOpen
  \bibfield  {author} {\bibinfo {author} {\bibfnamefont {J.}~\bibnamefont
  {Eun}}, \bibinfo {author} {\bibfnamefont {Z.}~\bibnamefont {Wang}}, \ and\
  \bibinfo {author} {\bibfnamefont {S.}~\bibnamefont {Chakravarty}},\
  }\href@noop {} {\bibfield  {journal} {\bibinfo  {journal} {Proceedings of the
  National Academy of Sciences}\ }\textbf {\bibinfo {volume} {109}},\ \bibinfo
  {pages} {13198} (\bibinfo {year} {2012})}\BibitemShut {NoStop}%
\bibitem [{Note4()}]{Note4}%
  \BibitemOpen
  \bibinfo {note} {A recent pre-print considering quantum oscillations in
  single layers with incommensurate density wave order found very similar
  patterns of Fermi surface reconstruction.\cite
  {allais2014connecting}}\BibitemShut {NoStop}%
\bibitem [{\citenamefont {Harrison}\ and\ \citenamefont
  {Sebastian}(2011)}]{harrison2011protected}%
  \BibitemOpen
  \bibfield  {author} {\bibinfo {author} {\bibfnamefont {N.}~\bibnamefont
  {Harrison}}\ and\ \bibinfo {author} {\bibfnamefont {S.}~\bibnamefont
  {Sebastian}},\ }\href@noop {} {\bibfield  {journal} {\bibinfo  {journal}
  {Physical Review Letters}\ }\textbf {\bibinfo {volume} {106}},\ \bibinfo
  {pages} {226402} (\bibinfo {year} {2011})}\BibitemShut {NoStop}%
\bibitem [{\citenamefont {Sebastian}\ \emph {et~al.}(2014)\citenamefont
  {Sebastian}, \citenamefont {Harrison}, \citenamefont {Balakirev},
  \citenamefont {Altarawneh}, \citenamefont {Goddard}, \citenamefont {Liang},
  \citenamefont {Bonn}, \citenamefont {Hardy},\ and\ \citenamefont
  {Lonzarich}}]{sebastian2014normal}%
  \BibitemOpen
  \bibfield  {author} {\bibinfo {author} {\bibfnamefont {S.~E.}\ \bibnamefont
  {Sebastian}}, \bibinfo {author} {\bibfnamefont {N.}~\bibnamefont {Harrison}},
  \bibinfo {author} {\bibfnamefont {F.}~\bibnamefont {Balakirev}}, \bibinfo
  {author} {\bibfnamefont {M.}~\bibnamefont {Altarawneh}}, \bibinfo {author}
  {\bibfnamefont {P.}~\bibnamefont {Goddard}}, \bibinfo {author} {\bibfnamefont
  {R.}~\bibnamefont {Liang}}, \bibinfo {author} {\bibfnamefont
  {D.}~\bibnamefont {Bonn}}, \bibinfo {author} {\bibfnamefont {W.}~\bibnamefont
  {Hardy}}, \ and\ \bibinfo {author} {\bibfnamefont {G.}~\bibnamefont
  {Lonzarich}},\ }\href@noop {} {\bibfield  {journal} {\bibinfo  {journal}
  {Nature}\ }\textbf {\bibinfo {volume} {511}},\ \bibinfo {pages} {61}
  (\bibinfo {year} {2014})}\BibitemShut {NoStop}%
\bibitem [{\citenamefont {Imry}\ and\ \citenamefont {Ma}(1975)}]{ImryMa}%
  \BibitemOpen
  \bibfield  {author} {\bibinfo {author} {\bibfnamefont {Y.}~\bibnamefont
  {Imry}}\ and\ \bibinfo {author} {\bibfnamefont {S.-k.}\ \bibnamefont {Ma}},\
  }\href@noop {} {\bibfield  {journal} {\bibinfo  {journal} {Physical Review
  Letters}\ }\textbf {\bibinfo {volume} {35}},\ \bibinfo {pages} {1399}
  (\bibinfo {year} {1975})}\BibitemShut {NoStop}%
\bibitem [{\citenamefont {Lawler}\ \emph {et~al.}(2010)\citenamefont {Lawler},
  \citenamefont {Fujita}, \citenamefont {Lee}, \citenamefont {Schmidt},
  \citenamefont {Kohsaka}, \citenamefont {Kim}, \citenamefont {Eisaki},
  \citenamefont {Uchida}, \citenamefont {Davis}, \citenamefont {Sethna} \emph
  {et~al.}}]{lawler2010intra}%
  \BibitemOpen
  \bibfield  {author} {\bibinfo {author} {\bibfnamefont {M.}~\bibnamefont
  {Lawler}}, \bibinfo {author} {\bibfnamefont {K.}~\bibnamefont {Fujita}},
  \bibinfo {author} {\bibfnamefont {J.}~\bibnamefont {Lee}}, \bibinfo {author}
  {\bibfnamefont {A.}~\bibnamefont {Schmidt}}, \bibinfo {author} {\bibfnamefont
  {Y.}~\bibnamefont {Kohsaka}}, \bibinfo {author} {\bibfnamefont {C.~K.}\
  \bibnamefont {Kim}}, \bibinfo {author} {\bibfnamefont {H.}~\bibnamefont
  {Eisaki}}, \bibinfo {author} {\bibfnamefont {S.}~\bibnamefont {Uchida}},
  \bibinfo {author} {\bibfnamefont {J.}~\bibnamefont {Davis}}, \bibinfo
  {author} {\bibfnamefont {J.}~\bibnamefont {Sethna}},  \emph {et~al.},\
  }\href@noop {} {\bibfield  {journal} {\bibinfo  {journal} {Nature}\ }\textbf
  {\bibinfo {volume} {466}},\ \bibinfo {pages} {347} (\bibinfo {year}
  {2010})}\BibitemShut {NoStop}%
\bibitem [{\citenamefont {Nie}\ \emph {et~al.}(2013)\citenamefont {Nie},
  \citenamefont {Tarjus},\ and\ \citenamefont {Kivelson}}]{nie2013quenched}%
  \BibitemOpen
  \bibfield  {author} {\bibinfo {author} {\bibfnamefont {L.}~\bibnamefont
  {Nie}}, \bibinfo {author} {\bibfnamefont {G.}~\bibnamefont {Tarjus}}, \ and\
  \bibinfo {author} {\bibfnamefont {S.}~\bibnamefont {Kivelson}},\ }\href@noop
  {} {\bibfield  {journal} {\bibinfo  {journal} {arXiv preprint
  arXiv:1311.5580}\ } (\bibinfo {year} {2013})}\BibitemShut {NoStop}%
\bibitem [{Note5()}]{Note5}%
  \BibitemOpen
  \bibinfo {note} {Under certain conditions, the criss-crossed phase may become
  chiral with the handedness of the chirality breaking a $Z_2$ symmetry. This
  is possible when bond density waves (with order parameter $\protect
  \mathaccentV {vec}17E{\protect \bm {\Delta }}_{\lambda }$) are also present
  in the clean system. A pattern of commensurate charge and bond density waves,
  which is perpendicular from layer to the next, but in which the CDW switches
  sign between alternate layers will be chiral, with the order parameter $\chi
  = \DOTSB \sum@ \slimits@ _{\lambda }\protect \bm {\protect \mathaccentV
  {hat}05E{z}}\cdot \left (\protect \bm {\protect \mathaccentV
  {hat}05E{n}}_{\lambda }\times \protect \bm {\protect \mathaccentV
  {hat}05E{n}}_{\lambda +1}\right )$, where the vector $\protect \bm
  {n}_{\lambda }$ in each layer is $\protect \bm {n}_{\lambda } = \phi
  _{\lambda ,x} \Delta ^{*}_{\lambda ,x}\protect \bm {\protect \mathaccentV
  {hat}05E{x}} + \phi _{\lambda ,y}\Delta ^{*}_{\lambda ,y}\protect \bm
  {\protect \mathaccentV {hat}05E{y}}$. This suggests a purely electronic
  mechanism for establishment of a chiral phase, unlike previous studies.\cite
  {jasper1,jasper2,jasper3}}\BibitemShut {NoStop}%
\bibitem [{\citenamefont {Li}\ \emph {et~al.}(2011)\citenamefont {Li},
  \citenamefont {Alidoust}, \citenamefont {Tranquada}, \citenamefont {Gu},\
  and\ \citenamefont {Ong}}]{NernstLBCO}%
  \BibitemOpen
  \bibfield  {author} {\bibinfo {author} {\bibfnamefont {L.}~\bibnamefont
  {Li}}, \bibinfo {author} {\bibfnamefont {N.}~\bibnamefont {Alidoust}},
  \bibinfo {author} {\bibfnamefont {J.~M.}\ \bibnamefont {Tranquada}}, \bibinfo
  {author} {\bibfnamefont {G.~D.}\ \bibnamefont {Gu}}, \ and\ \bibinfo {author}
  {\bibfnamefont {N.~P.}\ \bibnamefont {Ong}},\ }\href {\doibase
  10.1103/PhysRevLett.107.277001} {\bibfield  {journal} {\bibinfo  {journal}
  {Phys. Rev. Lett.}\ }\textbf {\bibinfo {volume} {107}},\ \bibinfo {pages}
  {277001} (\bibinfo {year} {2011})}\BibitemShut {NoStop}%
\bibitem [{\citenamefont {Lubashevsky}\ \emph {et~al.}(2014)\citenamefont
  {Lubashevsky}, \citenamefont {Pan}, \citenamefont {Kirzhner}, \citenamefont
  {Koren},\ and\ \citenamefont {Armitage}}]{PhysRevLett.112.147001}%
  \BibitemOpen
  \bibfield  {author} {\bibinfo {author} {\bibfnamefont {Y.}~\bibnamefont
  {Lubashevsky}}, \bibinfo {author} {\bibfnamefont {L.}~\bibnamefont {Pan}},
  \bibinfo {author} {\bibfnamefont {T.}~\bibnamefont {Kirzhner}}, \bibinfo
  {author} {\bibfnamefont {G.}~\bibnamefont {Koren}}, \ and\ \bibinfo {author}
  {\bibfnamefont {N.}~\bibnamefont {Armitage}},\ }\href {\doibase
  10.1103/PhysRevLett.112.147001} {\bibfield  {journal} {\bibinfo  {journal}
  {Phys. Rev. Lett.}\ }\textbf {\bibinfo {volume} {112}},\ \bibinfo {pages}
  {147001} (\bibinfo {year} {2014})}\BibitemShut {NoStop}%
\bibitem [{\citenamefont {Xia}\ \emph {et~al.}(2008)\citenamefont {Xia},
  \citenamefont {Schemm}, \citenamefont {Deutscher}, \citenamefont {Kivelson},
  \citenamefont {Bonn}, \citenamefont {Hardy}, \citenamefont {Liang},
  \citenamefont {Siemons}, \citenamefont {Koster}, \citenamefont {Fejer},\ and\
  \citenamefont {Kapitulnik}}]{KerrYBCO}%
  \BibitemOpen
  \bibfield  {author} {\bibinfo {author} {\bibfnamefont {J.}~\bibnamefont
  {Xia}}, \bibinfo {author} {\bibfnamefont {E.}~\bibnamefont {Schemm}},
  \bibinfo {author} {\bibfnamefont {G.}~\bibnamefont {Deutscher}}, \bibinfo
  {author} {\bibfnamefont {S.~A.}\ \bibnamefont {Kivelson}}, \bibinfo {author}
  {\bibfnamefont {D.~A.}\ \bibnamefont {Bonn}}, \bibinfo {author}
  {\bibfnamefont {W.~N.}\ \bibnamefont {Hardy}}, \bibinfo {author}
  {\bibfnamefont {R.}~\bibnamefont {Liang}}, \bibinfo {author} {\bibfnamefont
  {W.}~\bibnamefont {Siemons}}, \bibinfo {author} {\bibfnamefont
  {G.}~\bibnamefont {Koster}}, \bibinfo {author} {\bibfnamefont {M.~M.}\
  \bibnamefont {Fejer}}, \ and\ \bibinfo {author} {\bibfnamefont
  {A.}~\bibnamefont {Kapitulnik}},\ }\href {\doibase
  10.1103/PhysRevLett.100.127002} {\bibfield  {journal} {\bibinfo  {journal}
  {Phys. Rev. Lett.}\ }\textbf {\bibinfo {volume} {100}},\ \bibinfo {pages}
  {127002} (\bibinfo {year} {2008})}\BibitemShut {NoStop}%
\bibitem [{\citenamefont {He}\ \emph {et~al.}(2011)\citenamefont {He},
  \citenamefont {Hashimoto}, \citenamefont {Karapetyan}, \citenamefont
  {Koralek}, \citenamefont {Hinton}, \citenamefont {Testaud}, \citenamefont
  {Nathan}, \citenamefont {Yoshida}, \citenamefont {Yao}, \citenamefont
  {Tanaka} \emph {et~al.}}]{He2011}%
  \BibitemOpen
  \bibfield  {author} {\bibinfo {author} {\bibfnamefont {R.}~\bibnamefont
  {He}}, \bibinfo {author} {\bibfnamefont {M.}~\bibnamefont {Hashimoto}},
  \bibinfo {author} {\bibfnamefont {H.}~\bibnamefont {Karapetyan}}, \bibinfo
  {author} {\bibfnamefont {J.}~\bibnamefont {Koralek}}, \bibinfo {author}
  {\bibfnamefont {J.}~\bibnamefont {Hinton}}, \bibinfo {author} {\bibfnamefont
  {J.}~\bibnamefont {Testaud}}, \bibinfo {author} {\bibfnamefont
  {V.}~\bibnamefont {Nathan}}, \bibinfo {author} {\bibfnamefont
  {Y.}~\bibnamefont {Yoshida}}, \bibinfo {author} {\bibfnamefont
  {H.}~\bibnamefont {Yao}}, \bibinfo {author} {\bibfnamefont {K.}~\bibnamefont
  {Tanaka}},  \emph {et~al.},\ }\href@noop {} {\bibfield  {journal} {\bibinfo
  {journal} {Science}\ }\textbf {\bibinfo {volume} {331}},\ \bibinfo {pages}
  {1579} (\bibinfo {year} {2011})}\BibitemShut {NoStop}%
\bibitem [{\citenamefont {Karapetyan}\ \emph {et~al.}(2012)\citenamefont
  {Karapetyan}, \citenamefont {H\"ucker}, \citenamefont {Gu}, \citenamefont
  {Tranquada}, \citenamefont {Fejer}, \citenamefont {Xia},\ and\ \citenamefont
  {Kapitulnik}}]{KerrLBCO}%
  \BibitemOpen
  \bibfield  {author} {\bibinfo {author} {\bibfnamefont {H.}~\bibnamefont
  {Karapetyan}}, \bibinfo {author} {\bibfnamefont {M.}~\bibnamefont
  {H\"ucker}}, \bibinfo {author} {\bibfnamefont {G.~D.}\ \bibnamefont {Gu}},
  \bibinfo {author} {\bibfnamefont {J.~M.}\ \bibnamefont {Tranquada}}, \bibinfo
  {author} {\bibfnamefont {M.~M.}\ \bibnamefont {Fejer}}, \bibinfo {author}
  {\bibfnamefont {J.}~\bibnamefont {Xia}}, \ and\ \bibinfo {author}
  {\bibfnamefont {A.}~\bibnamefont {Kapitulnik}},\ }\href {\doibase
  10.1103/PhysRevLett.109.147001} {\bibfield  {journal} {\bibinfo  {journal}
  {Phys. Rev. Lett.}\ }\textbf {\bibinfo {volume} {109}},\ \bibinfo {pages}
  {147001} (\bibinfo {year} {2012})}\BibitemShut {NoStop}%
\bibitem [{\citenamefont {{Hosur}}\ \emph {et~al.}(2014)\citenamefont
  {{Hosur}}, \citenamefont {{Kapitulnik}}, \citenamefont {{Kivelson}},
  \citenamefont {{Orenstein}}, \citenamefont {{Raghu}}, \citenamefont {{Cho}},\
  and\ \citenamefont {{Fried}}}]{Hosur2}%
  \BibitemOpen
  \bibfield  {author} {\bibinfo {author} {\bibfnamefont {P.}~\bibnamefont
  {{Hosur}}}, \bibinfo {author} {\bibfnamefont {A.}~\bibnamefont
  {{Kapitulnik}}}, \bibinfo {author} {\bibfnamefont {S.~A.}\ \bibnamefont
  {{Kivelson}}}, \bibinfo {author} {\bibfnamefont {J.}~\bibnamefont
  {{Orenstein}}}, \bibinfo {author} {\bibfnamefont {S.}~\bibnamefont
  {{Raghu}}}, \bibinfo {author} {\bibfnamefont {W.}~\bibnamefont {{Cho}}}, \
  and\ \bibinfo {author} {\bibfnamefont {A.}~\bibnamefont {{Fried}}},\
  }\href@noop {} {\bibfield  {journal} {\bibinfo  {journal} {ArXiv e-prints}\ }
  (\bibinfo {year} {2014})},\ \Eprint {http://arxiv.org/abs/1405.0752}
  {arXiv:1405.0752 [cond-mat.supr-con]} \BibitemShut {NoStop}%
\bibitem [{\citenamefont {Robertson}\ \emph {et~al.}(2006)\citenamefont
  {Robertson}, \citenamefont {Kivelson}, \citenamefont {Fradkin}, \citenamefont
  {Fang},\ and\ \citenamefont {Kapitulnik}}]{robertson2006distinguishing}%
  \BibitemOpen
  \bibfield  {author} {\bibinfo {author} {\bibfnamefont {J.~A.}\ \bibnamefont
  {Robertson}}, \bibinfo {author} {\bibfnamefont {S.~A.}\ \bibnamefont
  {Kivelson}}, \bibinfo {author} {\bibfnamefont {E.}~\bibnamefont {Fradkin}},
  \bibinfo {author} {\bibfnamefont {A.~C.}\ \bibnamefont {Fang}}, \ and\
  \bibinfo {author} {\bibfnamefont {A.}~\bibnamefont {Kapitulnik}},\
  }\href@noop {} {\bibfield  {journal} {\bibinfo  {journal} {Physical Review
  B}\ }\textbf {\bibinfo {volume} {74}},\ \bibinfo {pages} {134507} (\bibinfo
  {year} {2006})}\BibitemShut {NoStop}%
\bibitem [{\citenamefont {Del~Maestro}\ and\ \citenamefont
  {Sachdev}(2005)}]{DelMaestro1}%
  \BibitemOpen
  \bibfield  {author} {\bibinfo {author} {\bibfnamefont {A.}~\bibnamefont
  {Del~Maestro}}\ and\ \bibinfo {author} {\bibfnamefont {S.}~\bibnamefont
  {Sachdev}},\ }\href {\doibase 10.1103/PhysRevB.71.184511} {\bibfield
  {journal} {\bibinfo  {journal} {Phys. Rev. B}\ }\textbf {\bibinfo {volume}
  {71}},\ \bibinfo {pages} {184511} (\bibinfo {year} {2005})}\BibitemShut
  {NoStop}%
\bibitem [{\citenamefont {Del~Maestro}\ \emph {et~al.}(2006)\citenamefont
  {Del~Maestro}, \citenamefont {Rosenow},\ and\ \citenamefont
  {Sachdev}}]{DelMaestro2}%
  \BibitemOpen
  \bibfield  {author} {\bibinfo {author} {\bibfnamefont {A.}~\bibnamefont
  {Del~Maestro}}, \bibinfo {author} {\bibfnamefont {B.}~\bibnamefont
  {Rosenow}}, \ and\ \bibinfo {author} {\bibfnamefont {S.}~\bibnamefont
  {Sachdev}},\ }\href {\doibase 10.1103/PhysRevB.74.024520} {\bibfield
  {journal} {\bibinfo  {journal} {Phys. Rev. B}\ }\textbf {\bibinfo {volume}
  {74}},\ \bibinfo {pages} {024520} (\bibinfo {year} {2006})}\BibitemShut
  {NoStop}%
\bibitem [{\citenamefont {Allais}\ \emph {et~al.}(2014)\citenamefont {Allais},
  \citenamefont {Chowdhury},\ and\ \citenamefont
  {Sachdev}}]{allais2014connecting}%
  \BibitemOpen
  \bibfield  {author} {\bibinfo {author} {\bibfnamefont {A.}~\bibnamefont
  {Allais}}, \bibinfo {author} {\bibfnamefont {D.}~\bibnamefont {Chowdhury}}, \
  and\ \bibinfo {author} {\bibfnamefont {S.}~\bibnamefont {Sachdev}},\
  }\href@noop {} {\bibfield  {journal} {\bibinfo  {journal} {arXiv preprint
  arXiv:1406.0503}\ } (\bibinfo {year} {2014})}\BibitemShut {NoStop}%
\bibitem [{\citenamefont {van Wezel}(2011)}]{jasper1}%
  \BibitemOpen
  \bibfield  {author} {\bibinfo {author} {\bibfnamefont {J.}~\bibnamefont {van
  Wezel}},\ }\href@noop {} {\bibfield  {journal} {\bibinfo  {journal} {EPL
  (Europhysics Letters)}\ }\textbf {\bibinfo {volume} {96}},\ \bibinfo {pages}
  {67011} (\bibinfo {year} {2011})}\BibitemShut {NoStop}%
\bibitem [{\citenamefont {van Wezel}\ and\ \citenamefont
  {Littlewood}(2010)}]{jasper2}%
  \BibitemOpen
  \bibfield  {author} {\bibinfo {author} {\bibfnamefont {J.}~\bibnamefont {van
  Wezel}}\ and\ \bibinfo {author} {\bibfnamefont {P.}~\bibnamefont
  {Littlewood}},\ }\href@noop {} {\bibfield  {journal} {\bibinfo  {journal}
  {Physcs Online Journal}\ }\textbf {\bibinfo {volume} {3}},\ \bibinfo {pages}
  {87} (\bibinfo {year} {2010})}\BibitemShut {NoStop}%
\bibitem [{\citenamefont {Castellan}\ \emph {et~al.}(2013)\citenamefont
  {Castellan}, \citenamefont {Rosenkranz}, \citenamefont {Osborn},
  \citenamefont {Li}, \citenamefont {Gray}, \citenamefont {Luo}, \citenamefont
  {Welp}, \citenamefont {Karapetrov}, \citenamefont {Ruff},\ and\ \citenamefont
  {van Wezel}}]{jasper3}%
  \BibitemOpen
  \bibfield  {author} {\bibinfo {author} {\bibfnamefont {J.-P.}\ \bibnamefont
  {Castellan}}, \bibinfo {author} {\bibfnamefont {S.}~\bibnamefont
  {Rosenkranz}}, \bibinfo {author} {\bibfnamefont {R.}~\bibnamefont {Osborn}},
  \bibinfo {author} {\bibfnamefont {Q.}~\bibnamefont {Li}}, \bibinfo {author}
  {\bibfnamefont {K.}~\bibnamefont {Gray}}, \bibinfo {author} {\bibfnamefont
  {X.}~\bibnamefont {Luo}}, \bibinfo {author} {\bibfnamefont {U.}~\bibnamefont
  {Welp}}, \bibinfo {author} {\bibfnamefont {G.}~\bibnamefont {Karapetrov}},
  \bibinfo {author} {\bibfnamefont {J.}~\bibnamefont {Ruff}}, \ and\ \bibinfo
  {author} {\bibfnamefont {J.}~\bibnamefont {van Wezel}},\ }\href@noop {}
  {\bibfield  {journal} {\bibinfo  {journal} {Physical Review Letters}\
  }\textbf {\bibinfo {volume} {110}},\ \bibinfo {pages} {196404} (\bibinfo
  {year} {2013})}\BibitemShut {NoStop}%
\end{thebibliography}%

\pagebreak
\widetext
\appendix

%----------------------------------------------------------------------------------------------------------------------------------------------------

\section{Landau Ginzburg theory}
The Landau Ginzburg theory for this model of two layers can be introduced by starting with a Hamiltonian that produces density waves within each layer at wave-vector $\Q_i$. This takes the form
\begin{align}
H = \sum_{\k} \sum_{\lambda,\lambda^\prime} (\varepsilon_{\k;\lambda\lambda^\prime}-\mu\delta_{\lambda\lambda^{\prime}}) \left(\cdag_{\k\lambda}\c_{\k,\lambda^\prime} + h.c.\right) + \sum_{\k,\kp;i}\sum_{\lambda,\lambda^{\prime}}V_{\lambda,\lambda^{\prime}}(\Q_{i}) \cdag_{\k+\Q_i,\lambda}\c_{\k,\lambda} \cdag_{\kp-\Q_i,\lambda}\c_{\kp,\lambda}.
\end{align}
where the interaction matrix is $V_{\lambda,\lambda^{\prime}} = -V\delta_{\lambda\lambda^{\prime}} + (V_{\perp}/2)\tau^{x}_{\lambda\lambda^{\prime}}$, while the kinetic term is $\varepsilon_{\k;\lambda\lambda^\prime} =\varepsilon(\k)\delta_{\lambda\lambda^{\prime}} + t_{\perp}(\k)\tau^{x}_{\lambda\lambda^\prime}$ . The partition function then takes the form
\begin{align}
Z &= \int \mathcal{D}\left[\psibar,\psi\right] e^{-S[\psibar,\psi]}\nonumber\\
S &= \int^{1/T}_{0}d\tau \mathcal{L}(k,\tau)
\end{align}
where the Lagrangian is given by 
\begin{align}
\mathcal{L} &= \sum_{\k,\lambda,\lambda^\prime} \psibar_{\k,\lambda} \left(\partial_{\tau}\delta_{\lambda\lambda^\prime} + \varepsilon_{\k;\lambda\lambda^\prime} -\mu \right)\psi_{\k,\lambda^\prime} - \sum_{\k,\k^\prime,i}\sum_{\lambda,\lambda^{\prime}}V_{\lambda,\lambda^{\prime}} \psibar_{\k+\Q_i,\lambda}\psi_{\k,\lambda} \psibar_{\kp-\Q_i,\lambda}\psi_{\kp,\lambda}.
\end{align}
We then perform the usual Hubbard-Stratonovich decomposition, by introducing the auxiliary fields $\phi_{\lambda}(\Q_i) = \sum_{\k}\psibar_{\kp-\Q_i,\lambda}\psi_{\kp,\lambda}$ (no sum over $\lambda$).  The resulting Lagrangian has the form
\begin{align}
\mathcal{L} = \sum_{i;\lambda,\lambda^{\prime}}V^{-1}_{\lambda,\lambda^\prime}\phi_{\lambda}(\Q_i)\phi^{*}_{\lambda^{\prime}}(\Q_i) + \sum_{\k,\k^\prime;\lambda,\lambda^\prime} \psibar_{\k\lambda}\mathcal{G}^{-1}_{\k,\kp;\lambda,\lambda^{\prime}} \psi_{\k^\prime,\lambda^\prime}
\end{align}
where 
\begin{align}
V^{-1}_{\lambda,\lambda^{\prime}}  = -\frac{1}{V}\delta_{\lambda,\lambda^{\prime}} + \frac{V_{\perp}}{2V^2}\tau^{x}_{\lambda,\lambda^{\prime}} + \mathcal{O}\left(\frac{V^{2}_{\perp}}{V^3}\right)
\end{align}
We can now integrate out fermions exactly, to obtain a Lagrangian purely in terms of the fields $\phi_{\lambda}(\Q_i)=\phi_{\lambda,i}$. The new Lagrangian is 
\begin{align}
\mathcal{L} =   \sum_{i;\lambda,\lambda^{\prime}}V^{-1}_{\lambda,\lambda^\prime}\phi_{\lambda,i}\phi^{*}_{\lambda^{\prime},i} - \text{Tr } \ln \mathcal{G}^{-1}(\phi_{\lambda,i})
\end{align}
Instead of focussing on specific $\Q_i's$, we can treat some general $\q$, in which case the full Green's function is given by 
\begin{align}
\mathcal{G}^{-1} = \left[-G^{-1}_{0}\delta_{k,\kpp} - \phi_{\lambda,\q}\delta_{\kpp,k-q}\right]\delta_{\lambda,\lambda^\prime} + t_{\perp}(\k)\delta_{k,\kpp}\tau^{x}_{\lambda,\lambda^\prime},
\end{align}
where we have defined the bare Green's function
\begin{align}
G_{0}(k) = G_{0}(\k,i\omega_n) = \frac{1}{i\omega_n - \varepsilon_{\k}-\mu}
\end{align}
Upon performing this expansion, up to second order in $t_{\perp}$ and fourth order in $\phi$, we find that the free energy takes the form 
\begin{align}
\mathcal{L} &= r_1 \left( \lvert\phi_{1x}\rvert^2+ \lvert\phi_{1y}\rvert^2+ \lvert\phi_{2x}\rvert^2+ \lvert\phi_{2y}\rvert^2\right) + r_2 \left( \phi_{1x}\phi^{*}_{2x} +\phi_{1y}\phi^{*}_{2y}\right) + u_1 \left(\lvert\phi_{1x}\rvert^4+\lvert\phi_{1y}\rvert^4+\lvert\phi_{2x}\rvert^4+\lvert\phi_{2y}\rvert^4\right)\nonumber\\
&+ v_1\left(\lvert\phi_{1x}\phi^{*}_{1y}\rvert^2+\lvert\phi^{}_{2x}\phi^{*}_{2y}\rvert^2\right)+v_2\left(\lvert\phi^{}_{1x}\phi^{*}_{2x}\rvert^2+\lvert\phi^{}_{1y}\phi^{*}_{2y}\rvert^2\right) + v_4\left(\lvert\phi^{}_{1x}\phi^{*}_{2y}\rvert+\lvert\phi^{}_{1y}\phi^{*}_{2x}\rvert^2\right)\nonumber\\
&+w_1 \left(\phi_{1x}\phi^{}_{1y}\phi^{*}_{2x}\phi^{*}_{2y} + c.c.\right) + w_2\left( \phi_{1x}\phi^{*}_{1y}\phi^{*}_{2x}\phi^{}_{2y} + c.c.\right)
\end{align}
We are primarily interested in the terms which involve couplings between layers, and these can be re-written in exactly the way we did in Eq.~\ref{eq:LGinterlayer}
\begin{align}
\mathcal{L}_{\text{interlayer}} &= \alpha\left( \phi^{}_{1,x}\phi^{*}_{2,x} + \phi^{}_{1,y}\phi^{*}_{2,y} + c.c.\right) +\beta\left\lvert  \phi^{}_{1,x}\phi^{*}_{2,x} + \phi^{}_{1,y}\phi^{*}_{2,y}\right\rvert^2 + \gamma\left\lvert \phi^{}_{1,x}\phi^{}_{2,y} - \phi^{}_{1,y}\phi^{}_{2,x}\right\rvert^2
\end{align}
\begin{figure*}
\begin{center}
        \includegraphics[width=0.35\textwidth]{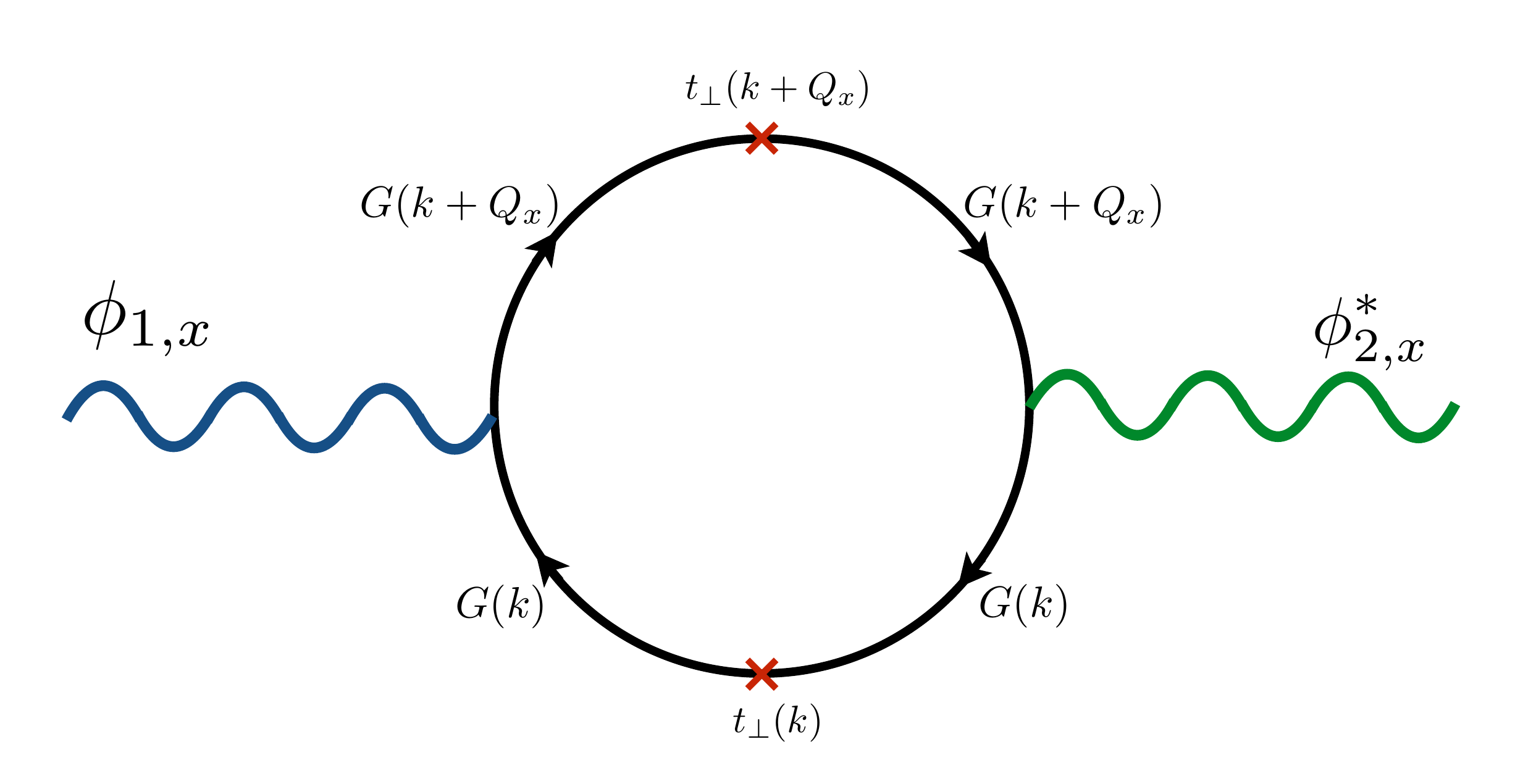}
              \caption{The Feynman diagram which contributes to the $\phi_{1x}\phi^{*}_{2x}$ term in the free energy expansion. }
               \label{fig:feyn1}
\end{center}
\end{figure*}

\begin{figure}
\begin{center}
        \subfloat[{The four diagrams for $\beta$ }]{
        \includegraphics[width=0.48\textwidth]{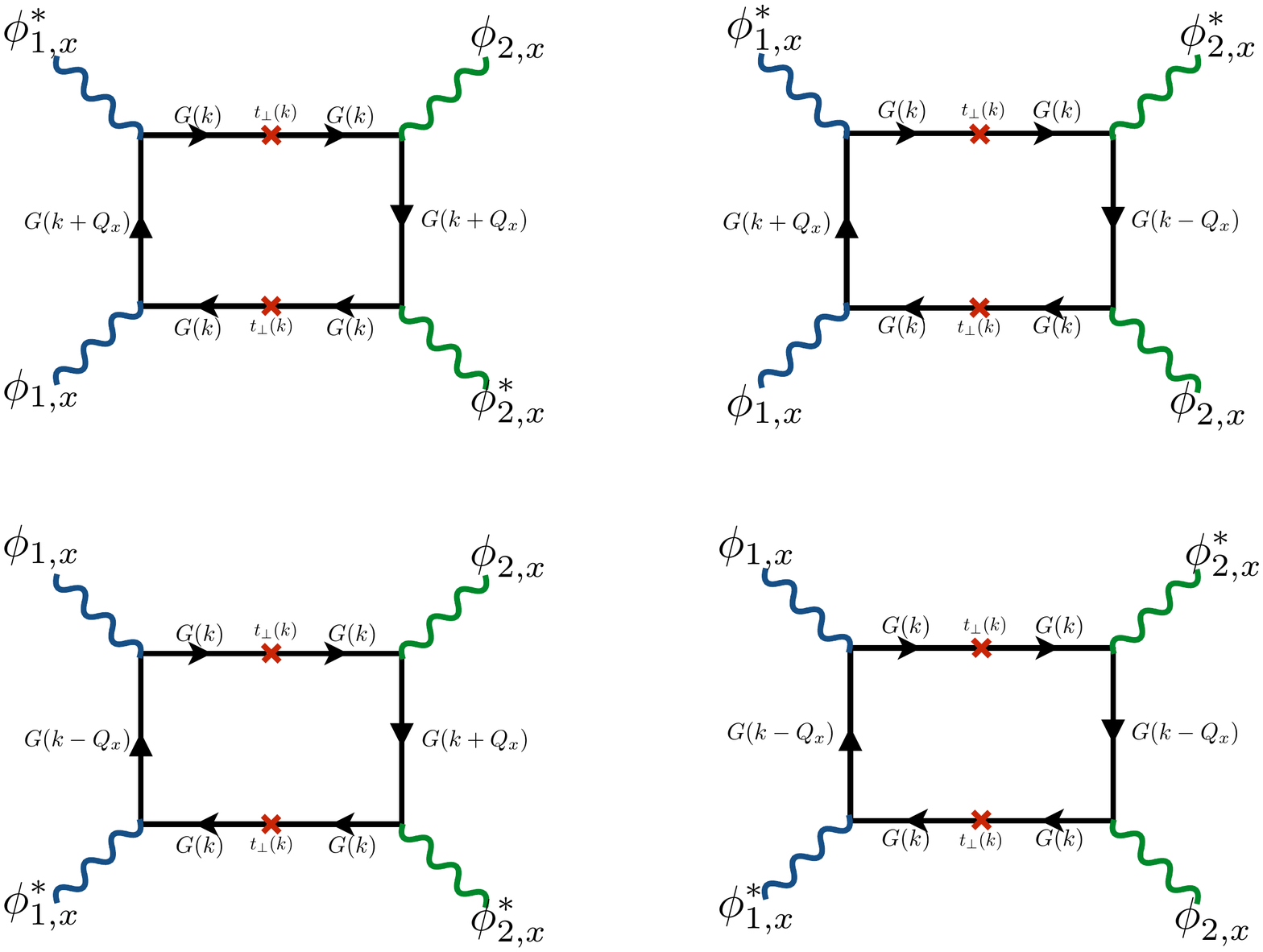}\label{fig:feynman1}}
        \quad
        \subfloat[{The four diagrams for $\gamma$ }]{
        \includegraphics[width=0.48\textwidth]{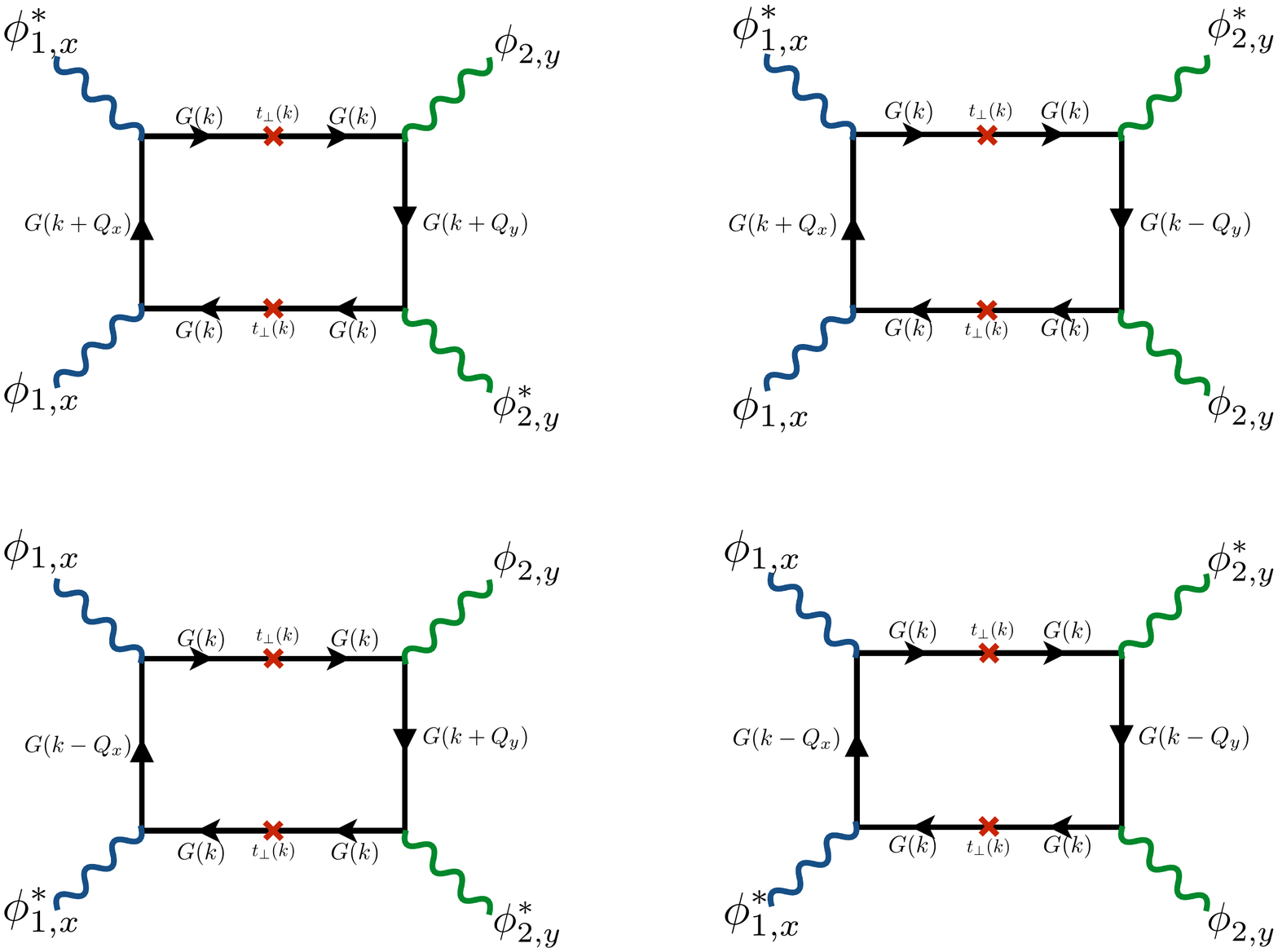}\label{fig:feynman2}}
       \caption{Feynman diagrams of quartic coefficients }
\end{center}
\end{figure}
where the coefficients are given by:
\begin{align}
\alpha &= \frac{V_{\perp}}{2V^2} + T\sum_{i\omega_n} \int \frac{d^2 k}{(2\pi)^2}\,\, G^2(k) G^2(k+q_x)t_{\perp}(\k)t_{\perp}(\k+\q_x) \\
\beta&= T \sum_{i\omega_n} \int \frac{d^2 k}{(2\pi)^2}\,\, G^4(k) \left[ G(k+q_x) + G(k-q_x)\right]^2t^{2}_{\perp}(\k) \label{eq:beta}\\
\gamma &= T \sum_{i\omega_n} \int \frac{d^2 k}{(2\pi)^2}\,\, G^4(k) \left[ G(k+q_x)+ G(k-q_x)\right]\left[G(k+q_y)+ G(k-q_y)\right] t^{2}_{\perp}(\k) \label{eq:gamma}
\end{align}
where $T$ is the temperature. These diagram for $\alpha$ is given in Fig.~\ref{fig:feyn1} while those for $\beta$ and $\gamma$ are shown in Figures~\ref{fig:feynman1} and~\ref{fig:feynman2}. 

\section{Analysis of free energy}
Before proceeding with the detailed evaluations of the coefficients in the Landau Ginzburg theory, let us examine its structure. We have divided the free energy into \textit{intra-} and \textit{inter-} layer terms:
$
F = F_{in\,plane} + F_{inter\,plane}, 
$ with
\begin{align}
F_{in\,plane} &=r\left( |\phi_{\lambda,x}|^2 + |\phi_{\lambda,y}|^2\right) +  u\left( |\phi_{\lambda,x}|^2 + |\phi_{\lambda,y}|^2\right)^2 + w|\phi_{\lambda,x}|^2|\phi_{\lambda,y}|^2.
\end{align}
The charge density waves onset when $r= (T-T_{cdw})<0$, and we will henceforth assume that $w>0$ so that a striped state is formed.
We have assumed that the dominant energy scales of the problem occur within each plane (interlayer terms come at $\mathcal{O}(t^{2}_{\perp})$), and so the magnitude of $\phi$ is controlled by these \textit{intra-}layer terms. Minimizing the above energy we have $|\phi_x| = \sqrt{|r|/2u}$ in the ordered state, where we have assumed that $\phi_y$ does not order. Then, using the vector notation $\vec{\phi}_{\lambda} = (\phi_{\lambda,x},\phi_{\lambda,y})$, the {\it inter}-layer free energy has the form:
\begin{align}
F_{inter-plane} = \alpha\left(\vec{\phi}_{1}\cdot\vec{\phi}^{*}_{2} + c.c.\right) + \beta\bigg\lvert\vec{\phi}_{1}\cdot\vec{\phi}^{*}_{2}\bigg\rvert^2 + \gamma\bigg\lvert \vec{\phi}_{1} \times \vec{\phi}_{2}\bigg\rvert^2
\label{eq:interlayer}.
\end{align}
This equation controls the orientation of stripes. Near to the CDW transition, we see that the bilinear coupling $\alpha$ must dominate and so forces parallel density waves. These can be either in-phase or out-of-phase depending on the sign of $\alpha$. Then, at lower temperature (as the magnitude of $\phi$ grows) there can be a first order transition from a parallel to a perpendicular phase provided $\gamma < \beta$. 

Schematically, we find that $\alpha$ is suppressed strongly by an interlayer tunneling that is momentum dependent, while $\eta \equiv \beta-\gamma$ is less affected because it contains contributions from many different parts of the Fermi surface. We also generically find that $\eta >0$, which can be seen from the form of 
Equations~\ref{eq:beta} and ~\ref{eq:gamma}. For general ordering vector $\Q$, we see that $\beta$ contains terms such as  $G^{4}(k)G^2(k\pm Q)$ which can connect pairs of hotspots (points on the Fermi surface), unlike the terms in $\gamma$ which involve momenta at three points, which are not generically located on the Fermi surface. We therefore find that $\beta$ is more singular than $\gamma$, and $\eta \equiv \beta -\gamma >0$.

Within a hotspot approximation, we find that for generic ordering vector magnitudes $|\Q|$ the coefficients are 
\begin{align}
\alpha &=  \frac{V_{\perp}}{2V^2} + \frac{t^{2}_{\perp \bm{1}}}{4\pi^2 v_x v_{y}}\frac{1}{v_y \Lambda}\label{eq:alpha1}\\
\beta &= \frac{t^2_{\perp \bm{1}}}{24\pi^2v_xv_y}\frac{1}{v^{3}_y\Lambda^{3}}\left[ 1 - 2\left(\frac{ v_y\Lambda}{\xi_{1-X}}\right)^2\right] + \frac{t^2_{\perp \bm{2}}}{\pi^2u_xu_y}\frac{1}{\left(u_x\Lambda\right)^3}\left(\frac{u_x\Lambda}{\xi_{3-X}}\right)^4 \label{eq:beta1}\\
\gamma &=-\frac{t^2_{\perp \bm{1}}}{12\pi^2 v_x v_y}\frac{1}{v^3_y \Lambda^3}\left[\left(\frac{v_y\Lambda}{\varepsilon_{1+Y}}\right)^2 +\left(\frac{v_y\Lambda}{\xi_{1-Y}}\right)^2 \right] + \frac{t^2_{\perp \bm{2}}}{\pi^2u_xu_y}\frac{1}{\left(u_x\Lambda\right)^3}\left[ \left(\frac{u_x\Lambda}{\xi_{3+Y}}\right)^4 +  \left(\frac{u_x\Lambda}{\xi_{3-Y}}\right)^4\right]\label{eq:gamma1}
\end{align}
where $v_x  \gg v_y$ and $u_x\sim v_y \ll u_y \sim v_x$ are Fermi velocities defined at the hotspots, $\Lambda$ is a momentum cutoff, and $\xi_{n\pm I}$ are constant ($\k$ independent) values of the dispersion at various non hotspot positions in the Brillouin zone (see Sec.~\ref{sec:setup} for details). The hotspot positions depend implicitly on $|\Q|$, so the dispersions $\vec{v}$ and $\vec{u}$ will also be $|\Q|$ dependent. We have assumed that $(v\Lambda/\xi) \ll 1$ which is true for generic (but not all!) momenta. The tunneling terms are equal for uniform interlayer tunneling: $t_{\perp \bm{1}} = t_{\perp \bm{2}} = t_{\perp}$, while for an interlayer tunneling that is momentum dependent they are generically less than unity (See Eq.~\ref{eq:tunnelingterm} for analytic expression).

\noindent Expanding in $\delta = (v_y\Lambda/\xi)\approx  (u_x\Lambda/\xi) \ll 1$, we find that
\begin{align}
\eta= \beta -\gamma \sim \frac{t^2_{\perp \bm{1}}}{24\pi^2v_xv_y}\frac{1}{v^{3}_y\Lambda^{3}}\left[ 1 + \mathcal{O}\left(\delta^2\right)\right] > 0
\end{align}
which is generally positive (i.e.  terms in $\gamma$ are subleading corrections to the $(1/v_y\Lambda)^3$ term of $\beta$.) This means the transition to a perpendicular state is always possible provided $\phi$ is large enough. Meanwhile we see that while $\alpha$ has contributions from only one hotspot, $\eta = \beta - \gamma$ has contributions from both hotspots, and so is not suppressed as strongly as $\alpha$. Thus the transition to a perpendicular phase, which occurs at a temperature $T_{\perp} = T_{cdw} - u\alpha/(\beta - \gamma)$ becomes possible as $\alpha/(\beta -\gamma)$ becomes smaller. 
\FloatBarrier
\section{Hotspot approximations for integrals}
\begin{figure*}
\begin{center}
        \includegraphics[width=0.5\textwidth]{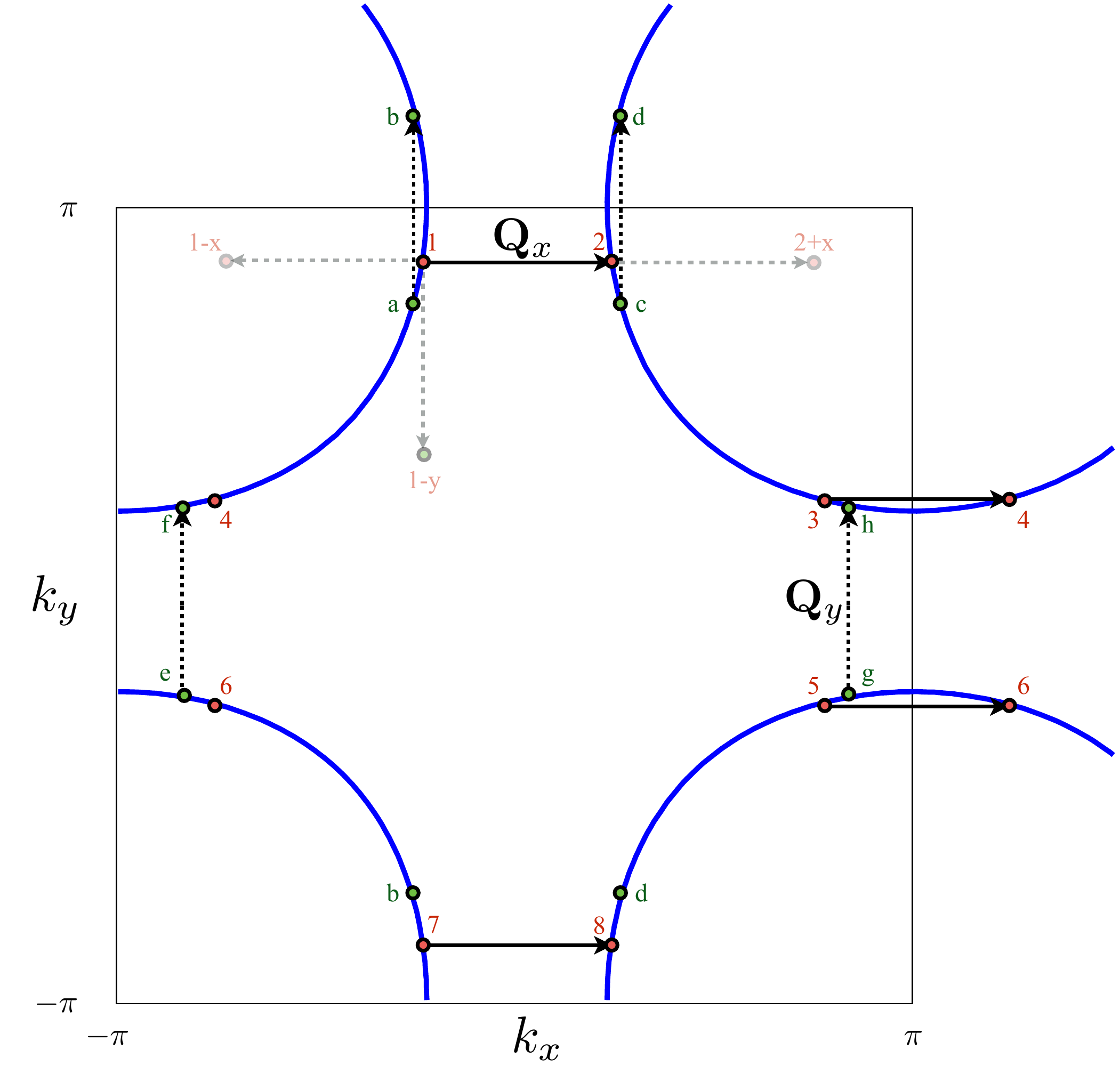}
              \caption{Hot spot topology }
               \label{fig:hss}
\end{center}
\end{figure*}
We can approximate the Landau Ginzburg coefficients by using hotspot approximations for the integrals. Here we schematically demonstrate what these coefficients are in terms of Green's functions defined at the hotspots:
\begin{align}
\alpha &=  \int_{\k,\omega_n}\,\, G^2(k) G^2(k+q_x)t_{\perp}(\k)t_{\perp}(\k+\q_x) \nonumber\\
&=\int_{\k,\omega_n} G^{2}_{1}G^{2}_{2}t_1t_2 + G^2_3 G^2_4t_3t_4 + G^2_5G^2_6t_5t_6 + G^2_7G^2_8t_7t_8 \nonumber\\
&=2\int_{\k,\omega_n} G^2_1G^2_2t^2_1 + G^2_3G^2_4t^2_3\\
&=2\left(t^{2}_1I_1 + t^{2}_3I_2\right)
\end{align}
where $G_{1} = G(\k_1 + k_x + k_y,i\omega_n) =\left(i\omega_n -\varepsilon_{\k_1 + k_x + k_y} - \mu\right)^{-1}$ is the Green's function defined near to hotspot point $1$ etc. Note that we have used the symmetry properties of $t_{\perp}(\k)$ (i.e. $t_1 = t_2$ etc.). 
\begin{figure}[b]
\begin{center}
        \includegraphics[width=0.45\textwidth]{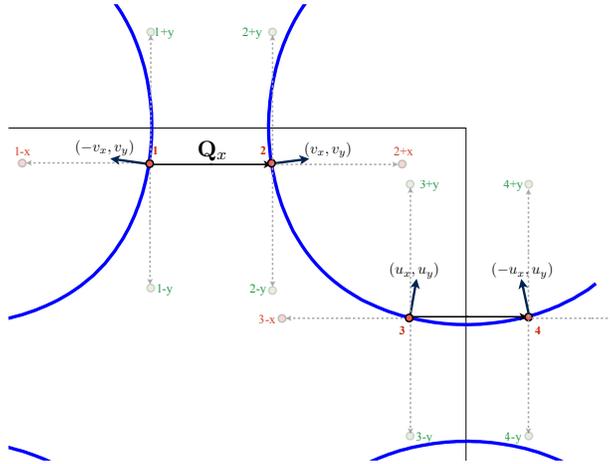}
              \caption{Dispersions in the vicinity of the hotspots}
               \label{fig:hsdets}
\end{center}
\end{figure}
For $\beta$, we have many more diagrams to consider and so we will henceforth drop the factors of $t_{\perp}$ for clarity (they can easily be restored as multiplicative factors within the hotspot approximation). For $\beta$ we have
\begin{align}
\beta&= \int_{\k,\omega_n} G^4(k)G^2(k+Q)+ G^4(k)G^2(k-Q)+ 2G^4(k)G(k+Q)G(k-Q)\nonumber\\
&=\beta_1 + \beta_2 + 2\beta_3
\end{align}
We see that while the first two terms above involve hotspots, the third does not generically connect three hotspots. We write Green's functions at these `cold-spots' by terms like $G_{1-x}$ to denote the spot displaced by distance  $-\Q_x$ from hotspot 1. With this sort of terminology, and using the $C_4$ and inversion symmetries of the band structure, we get
\begin{align}
\beta_1 &= 2\int_{\k,\omega_n} G^4_1G^2_2 + G^4_3G^2_4 = 2\left(I_3 + I_4\right)\\
\beta_2 &=2\int_{\k,\omega_n} G^4_2G^2_1 + G^4_4G^2_3= 2\left(I_5 + I_6\right) \\
\beta_3 &= 2\int_{\k,\omega_n} G^4_1G_2G_{1-x} + G^4_2G_1G_{2+x} +  G^4_3G_4G_{3-x} + G^4_4G_3G_{4+x} \nonumber\\
&= 4\int_{\k,\omega_n} G^4_1G_2G_{1-x}  +  G^4_3G_4G_{3-x} = 2\left(I_7 + I_8\right)
\end{align}
Finally for $\gamma$, we must now involve both the `red' and `green' hotspots in this calculation. All the integrals will involve at least one `cold' spot for generic values of $\Q$. We have 
\begin{align}
\gamma &=\int_{\k,\omega_n}G^4(k) \left[ G(k+Q_x)G(k+Q_y) + G(k+Q_x)G(k-Q_y) + G(k-Q_x)G(k+Q_y) + G(k-Q_x)G(k-Q_y)\right] t^{2}_{\perp}(\k) 
\end{align}
We set $\gamma_1 =\left(\gamma_1 + \gamma_2 + \gamma_3 + \gamma_4\right)$, so we can write each of these terms in the following way:
\begin{align}
\gamma_1 &= \int_{\k,\omega_n}G^4_1G_2G_{1+y} + G^4_3G_4G_{3+y} +G^4_5G_6G_{5+y} + G^{4}_7G_8G_{7+y} +  G^4_aG_{a+x}G_b + G^4_cG_{c+x}G_d + G^{4}_{g}G_{g+x}G_h + G^4_e G_{e+x}G_f \nonumber\\
 &=\int_{\k,\omega_n}G^4_1G_2G_{1+y} + G^4_3G_4G_{3+y} +G^4_5G_6G_{5+y} + G^{4}_7G_8G_{7+y} +  
G^{4}_3G_{3-y}G_4 + G^{4}_5G_{5-y}G_{6} + G^{4}_{7}G_{7-y}G_{8} + G^{4}_{1}G_{1-y}G_{2}\nonumber\\
&=\int_{\k,\omega_n}G^{4}_{1}G_{2}\left[G_{1+y} + G_{1-y}\right] + G^{4}_{3}G_{4}\left[G_{3+y} + G_{3-y}\right] + G^{4}_{5}G_{6}\left[G_{5+y} + G_{5-y}\right] + G^{4}_{7}G_8\left[G_{7+y}+G_{7-y}\right]\nonumber\\
&=2\int_{\k,\omega_n}G^{4}_{1}G_{2}\left[G_{1+y} + G_{1-y}\right] + G^{4}_{3}G_{4}\left[G_{3+y} + G_{3-y}\right]
\end{align}
Where each line exploits symmetries of the band structure. In fact, we can show that each of the terms $\gamma_1, \gamma_2, \gamma_3$ and $\gamma_4$ is equivalent under 90 degree rotation. They are all therefore equal to each other, and we have
\begin{align}
\gamma = 8\int_{\k,\omega_n}G^{4}_{1}G_{2}\left[G_{1+y} + G_{1-y}\right] + G^{4}_{3}G_{4}\left[G_{3+y} + G_{3-y}\right] =8\left[\left(I_9 + I_{10}\right) + \left(I_{11} + I_{12}\right)\right]\end{align}
We have therefore reduced the problem to one involving just 4 hotspots, and their related `cold spots.' We will then approximate the dispersions in the vicinity of these spots as shown in Fig.~\ref{fig:hsdets}

\section{Analytic evaluation of coefficients}
\subsection{General strategy}\label{sec:setup}

Within the hotspot approximation, we can find analytic expressions for the Landau Ginzburg coefficients. We will follow closely the work of \citet{wang2014charge} in evaluating these integrals. Note that we will choose the generic scenario where there are four pairs of hotspots for each ordering direction (See Fig.~\ref{fig:hss}). This occurs for $|\Q|>|\Q_c|$ where  
\begin{align*}
|\Q_{c}| = 2\cos^{-1}\left(\frac{2t-\mu}{2t+4t^{\prime}}\right).
\end{align*}
which ensures that $|\Q|$ is larger than the distance between Fermi surface points at the Brillouin zone edge. The four unique hotspots located at 
\begin{equation}
\begin{aligned}
\bm{k_1} &=\left(\frac{-Q_{x}}{2}, k_{y0}\right),\quad\qquad \bm{k_2} =\left(\frac{Q_{x}}{2}, k_{y0}\right),\nonumber\\
\bm{k_3} &=\left(\frac{\pi-Q_{x}}{2}, k_{y1}\right),\qquad\bm{k_4}=\left(\frac{-\pi+Q_{x}}{2}, k_{y1}\right),\nonumber\\
\end{aligned}
\end{equation}
where $k_{y0}$ is given by solving $-2t(\cos{Q_x} -\cos{k_{0}}) + 4t^{\prime}\cos{Q_{x}}\cos{k_{0}} -\mu = 0$, while $k_{y1}$ is given by a similar equation but with $Q_x/2$ replaced by $\pi-Q_x/2$. We have labeled these points by numbers $1-4$ as shown in Fig.~\ref{fig:hsdets}. We then linearize the dispersion in the vicinity of the first pair of these hotspots as 
\begin{align}
\quad \varepsilon_{\k_1}-\mu&\equiv \,\,\xi_1 \,\,\approx -v_{x}k_{x} + v_{y}k_{y} \\
\quad\varepsilon_{\k_2} -\mu &\equiv \,\,\xi_2 \,\,\approx v_{x}k_{x} + v_{y}k_{y}
\end{align}
with the approximation $v_x \gg v_y$, while at the second pars of hotspots, we have 
\begin{align}
\quad \varepsilon_{\k_3}-\mu&\equiv \,\,\xi_3\,\, \approx u_{x}k_{x} + u_{y}k_{y} \\
\quad\varepsilon_{\k_4} -\mu &\equiv\,\, \xi_4\,\, \approx -u_{x}k_{x} + u_{y}k_{y}
\end{align}
with $u_x \ll u_y$. Finally, for integrals involving `cold spots,' we write the dispersion as a constant in the vicinity of these spots, with terms like 
\begin{equation}
\varepsilon_{\k_1 - \Q_x} -\mu=\xi_{1-X}
\end{equation}
to denote the constant value at position $\bm{k_1} -\Q_x$ which does not lie on the Fermi surface (see Fig.~\ref{fig:hsdets})
The integrals are then done with momentum cutoffs $\Lambda$ around the hotspots, i.e. 
\begin{align}
\int\frac{d^{2}k}{(2\pi)^2} \approx  \frac{1}{4\pi^2} \int^{\Lambda}_{-\Lambda}dk_x\int^{\Lambda}_{-\Lambda}dk_y
\end{align}
and after performing the momentum integrals, we approximate the Matsubara sums as integrals:
\begin{align}
T\sum_{i\omega_n} \rightarrow \int^{\infty}_{|\omega|>\pi T}\frac{d\omega}{(2\pi)}
\end{align}
before taking the limit $T\rightarrow 0$. Note that many of the integrals are infra-red singular, so we keep temperature finite as a regulator of the infinity before seeing $T\rightarrow 0$ at the end of the integrals.\cite{wang2014charge}

\subsection{Evaluation of $\alpha$}
\begin{figure}[h!]
\begin{center}
  \includegraphics[width=0.48\textwidth]{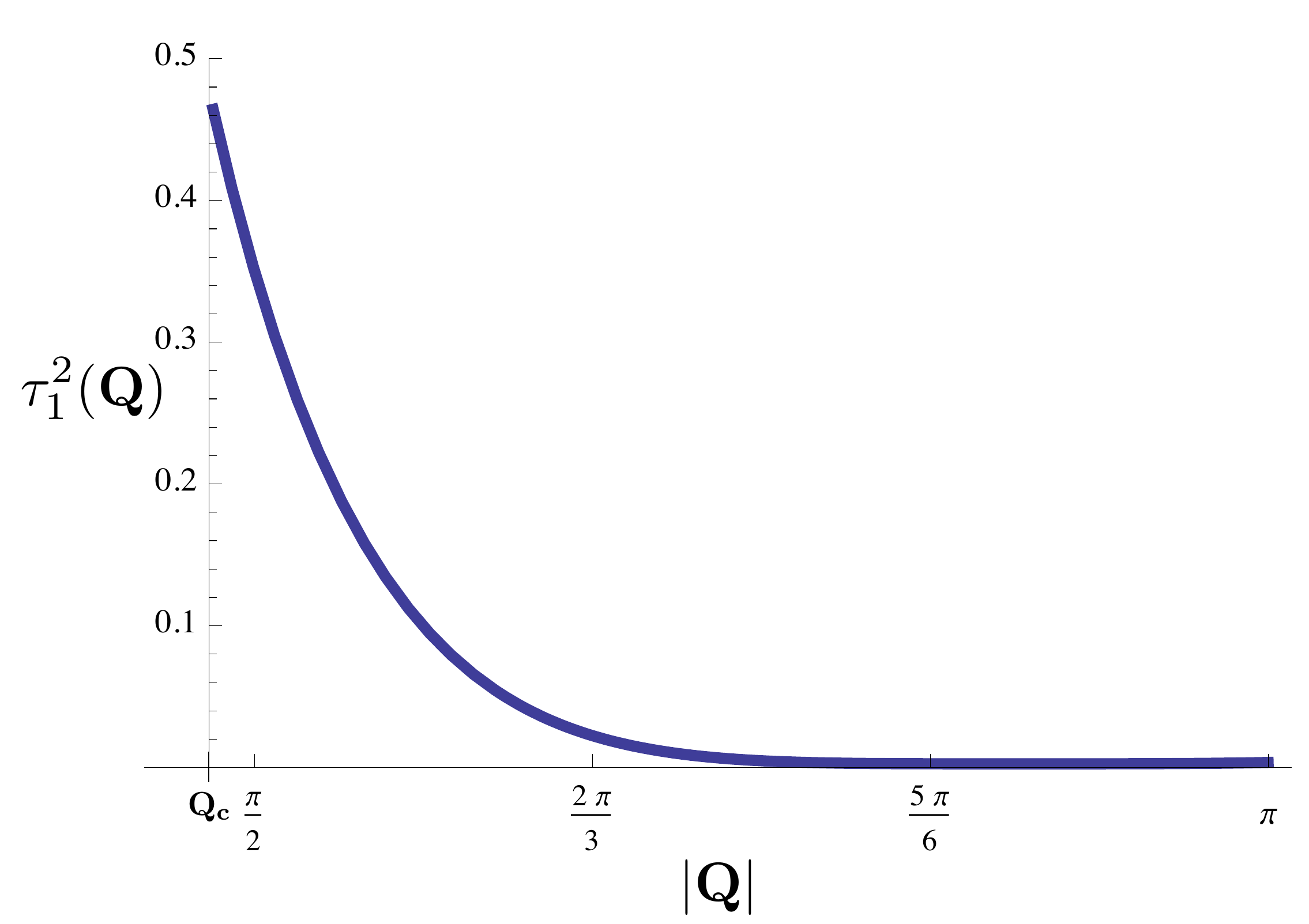}
       \caption{The ratio $\alpha_{\k-\text{dependent}}/\alpha_{\k-\text{independent}}$ which is proportional to the coefficient $\tau^{2}_{1}(\k)$ defined in Eq.~\ref{eq:suppresion1}. We see that the bilinear coupling is strongly suppressed by a $\k$ dependent interlayer tunneling. }
       \label{fig:supp}
\end{center}
\end{figure}

\noindent We have $\alpha = 2\left(t^{}_1t^{}_{2} I_1 + t^{}_3t^{}_4I_2\right)$ where
\begin{align}
t^{}_1t^{}_{2}I_1&=\int_{\k,\omega_n} G^{2}_{1}G^{2}_2 t^{}_1t^{}_2 \nonumber,\\
&=\frac{T}{4\pi^2}\sum_{i\omega_{n}} \int^{\Lambda}_{-\Lambda} dk_{x} \, \int^{\Lambda}_{-\Lambda_{}} dk_{y} \left[\frac{1}{i\omega_{n} -(-v_{x}k_{x} + v_{y}k_{y})}\right]^2\left[\frac{1}{i\omega_{n} -( v_{x}k_{x} +v_{y}k_{y})}\right]^2 t_{\perp}(\frac{-Q_x}{2},k_0)t_{\perp}(\frac{Q_x}{2},k_0)   \\
t^{}_3t^{}_{4}I_2&=\int_{\k,\omega_n} G^{2}_{3}G^{2}_4 t^{}_3t^{}_4 \nonumber\\
&=\frac{T}{4\pi^2}\sum_{i\omega_{n}} \int^{\Lambda_{}}_{-\Lambda_{}} dk_{x} \, \int^{\Lambda_{}}_{-\Lambda_{}} dk_{y} \left[\frac{1}{i\omega_{n} -(-u_{x}k_{x} + u_{y}k_{y})}\right]^2\left[\frac{1}{i\omega_{n} -( u_{x}k_{x} + u_{y}k_{y})}\right]^2 t_{\perp}(\pi-\frac{Q_x}{2},k_1)t_{\perp}(-\pi+\frac{Q_x}{2},k_1),\nonumber\\
\end{align}
where $\Lambda$ is the cutoff. We postpone an explicit evaluation of these integrals, and first examine how these integrals depend on the form of $t_{\perp}(\k)$. Since $t_{\perp}(\k)$ is even in $\k$, we see that 
\begin{align*}
 t_{\perp}(\frac{-Q_x}{2}+k_x,k_0+k_y)t_{\perp}(\frac{Q_x}{2}+k_x,k_0+k_y) \approx t^{2}_{\perp}(\frac{Q_x}{2},k_0) + \mathcal{O}(k_x,k_y).
 \end{align*}
 In the presence of uniform interlayer tunneling, $t_{\perp}(\k)=t_{\perp}$, so we have:
 \begin{align}
 \alpha_{\text{uniform}} = t^{2}_{\perp}\left(I_{1} + I_{2}\right).
 \end{align}
When the interlayer tunneling has the form $t_{\perp}(\k) = \frac{1}{4}t_{\perp}\left(\cos{k_x} -\cos{k_y}\right)^2$, then each hotspot integral comes with a prefactor, so that 
\begin{align}
\alpha_{\k-\text{dependent}} = t^{2}_{\perp}\bigg[ \tau^{2}_1(Q_x) I_{1}\,\, + \,\,\tau^2_2(Q_x) I_{2}\bigg],
\label{eq:suppresion1}
\end{align}
where the suppression prefactor for a cuprate like band structure of the form $\xi(\k) = -2t(\cos{k_x} + \cos{k_y}) + 4t^{\prime}\cos{k_x}\cos{k_y} -\mu$ is
\begin{align}
\tau_1(Q) = \frac{1}{4}\left[\frac{4t^{\prime}\cos^{2}{(Q/2)} -4t^{}\cos{(Q/2)} - \mu}{4t^{\prime}\cos{(Q/2)}-2t}\right]^2.
\label{eq:tunnelingterm}
\end{align}
While $\tau_1(Q)=\tau_2(2\pi-Q)$. These factors are both plotted in Fig.~\ref{fig:supp}. We therefore see that the interlayer tunneling introduces an overall multiplicative suppression factor in this hotspot approximation.

\noindent Using the linearized dispersion, the integral $I_1$ is then
\begin{equation}
I_1 = T\sum_{i\omega_n} \int^{\Lambda}_{-\Lambda} \frac{d^{2}k}{(2\pi)^2} \, \left[\frac{1}{i\omega_n + v_x k_x - v_y k_y}\right]^2\left[\frac{1}{i\omega_n - v_x k_x - v_y k_y}\right]^2.
\end{equation}
We will make the simplifying approximation $v_x \gg v_y$ here, which motivates the following change of variables:
\begin{equation}
x=v_x k_x, \quad y=v_yk_y, \quad \Lambda_x = v_x\Lambda, \quad\text{and    } \Lambda_y = v_y\Lambda_y,
\end{equation}
with $\Lambda_x \gg \Lambda_y$. With these new variables, the integral takes the form 
\begin{equation}
I_1 = \frac{T}{4\pi^2 v_x v_y} \sum_{i\omega_n} \int^{\Lambda_y}_{-\Lambda_y} dy \int^{\Lambda_x}_{-\Lambda_x} dx \left(\frac{1}{x+y-i\omega_n}\right)^2\left(\frac{1}{x-y+i\omega_n}\right)^2.
\end{equation}
Since $\Lambda_x \gg \Lambda_y$ we can separate the $x$-integral $\int^{\Lambda_x}_{-\Lambda_x} dx$ into $I_1 = \int^{\infty}_{-\infty} dx - \int_{\lvert x\rvert > \Lambda_x} dx = I_{1a} - I_{1b}$. So for $I_{1a}$ we get 
\begin{equation}
\begin{aligned}
I_{1a} &= \frac{T}{4\pi^2 v_x v_y} \sum_{i\omega_n} \int^{\Lambda_y}_{-\Lambda_y} dy \int^{\infty}_{-\infty} dx \left(\frac{1}{x+y-i\omega_n}\right)^2\left(\frac{1}{x-y+i\omega_n}\right)^2\\
&= \frac{iT}{8\pi v_x v_y}\sum_{i\omega_n} \text{sgn}(\omega_n) \int^{\Lambda_y}_{-\Lambda_y} dy \left(\frac{1}{y-i\omega_n}\right)^3\\
&=\frac{iT}{8\pi v_x v_y}\sum_{i\omega_n}  \int^{\Lambda_y}_{-\Lambda_y} dy \left(\frac{1}{y-i\lvert\omega_n\rvert}\right)^3\\
&=\frac{iT}{16\pi v_x v_y} \sum_{i\omega_n} \left[ \left(\frac{1}{\lvert \omega_n \rvert + i\Lambda_y}\right)^2 -\left(\frac{1}{\lvert \omega_n \rvert - i\Lambda_y}\right)^2 \right].
\end{aligned}
\end{equation}
We now assume that $\Lambda_y \gg T$ so that the Matsubara sum can be reasonably approximated by an integral as described earlier. With this approximation, the integral becomes:
\begin{equation}
\begin{aligned}
I_{1a}&\approx \frac{i}{32\pi^2 v_x v_y} \int d\omega \left[ \left(\frac{1}{\lvert \omega_n \rvert + i\Lambda_y}\right)^2 -\left(\frac{1}{\lvert \omega_n \rvert - i\Lambda_y}\right)^2 \right]\\
&= \frac{1}{8\pi^2 v_x v_y} \frac{1}{\Lambda_y} .
\end{aligned} 
\end{equation}
\noindent For the integral $I_{1b}$ we have 
\begin{equation}
I_{1b} = \frac{T}{4\pi^2 v_x v_y} \sum_{i\omega_n} \int^{\Lambda_y}_{-\Lambda_y} dy \int_{|x|>\Lambda_x} dx \left(\frac{1}{x+y-i\omega_n}\right)^2\left(\frac{1}{x-y+i\omega_n}\right)^2.
\end{equation}
By noting that the $\Lambda_x \gg \Lambda_y$, we can change the order of integration and Taylor expand the integrand. We then get 
\begin{equation}
\begin{aligned}
I_{1b} &\approx \frac{T\Lambda_y}{2\pi^2v_x v_y}\sum_{i\omega_n} \int_{|x|>\Lambda_x} dx \frac{1}{\left( x^2 + \omega_n^2\right)^2} \\
&=  \frac{T\Lambda_y}{\pi^2v_x v_y}\sum_{i\omega_n} \int^{\infty}_{\Lambda_x} dx \frac{1}{\left( x^2 + \omega_n^2\right)^2}\\
&= \frac{1}{8\pi^2 v_x v_y}\frac{ \Lambda_{y}}{\Lambda^{2}_{x}}.
\end{aligned}
\end{equation}
Combing the above two results, we find that 
\begin{equation}
\underline{I_{1} = I_{1a} -I_{1b} =\frac{1}{8\pi^2v_x v_y} \left[\frac{1}{\Lambda_y} - \frac{\Lambda_y}{\Lambda^{2}_{x}}\right] 
 \approx \frac{1}{8\pi^2v_x v^{2}_{y}}\frac{1}{\Lambda}.}
\end{equation}

We now turn to $I_{2}$. Recall that the hotspots are now labelled by velocity $\vec{u}$, where this time we have $u_{y}\gg u_{x}$. So we have 
\begin{equation}
I_2 = T\sum_{i\omega_n} \int^{\Lambda}_{-\Lambda} \frac{d^{2}k}{(2\pi)^2} \, \left[\frac{1}{i\omega_n + u_x k_x -u_y k_y}\right]^2\left[\frac{1}{i\omega_n - u_x k_x - u_y k_y}\right]^2,
\end{equation}
and we perform a similar change of variables:
\begin{equation}
x=u_x k_x, \quad y=u_yk_y, \quad \Lambda_x = u_x\Lambda, \quad\text{and    } \Lambda_y = u_y\Lambda_y,
\end{equation}
but now $\Lambda_y \gg \Lambda_x$. Under this change of variables, we get 
\begin{equation}
I_2 = \frac{T}{4\pi^2 u_x u_y} \sum_{i\omega_n} \int^{\Lambda_x}_{-\Lambda_x} dx \int^{\Lambda_y}_{-\Lambda_y} dy \left(\frac{1}{x+y-i\omega_n}\right)^2\left(\frac{1}{x-y+i\omega_n}\right)^2.
\end{equation}
We then do the same trick for the $y$ integral: $I_2= \int^{\infty}_{-\infty} dy - \int_{\lvert y\rvert > \Lambda_y} dy = I_{2a} - I_{2b}$. For $I_{2a}$ the $y$-integral vanishes since the poles are in the same half plane. The integral $I_{2b}$ also vanishes:
\begin{equation}
\begin{aligned} 
I_{2b} &= \frac{T}{4\pi^2 u_x u_y} \sum_{i\omega_n} \int^{\Lambda_x}_{-\Lambda_x} dx \int_{|y|>\Lambda_y} dy \,\left(\frac{1}{x+y-i\omega_n}\right)^2 \left(\frac{1}{x-y+i\omega_n}\right)^2\\
&\approx \frac{\Lambda_x}{8\pi^3 u_x u_y}\int d\omega\int^{\infty}_{\Lambda_y} dy  \left[ \left(\frac{1}{y-i\omega}\right)^4 + \left(\frac{1}{y+i\omega}\right)^4\right]\\
&=0.
\end{aligned}
\end{equation}

Thus we have the final result that the coefficient $\alpha$ is a cutoff dependent quantity that has contributions from only one pair of hotspots:
\begin{equation}
\boxed{
\alpha =  \frac{t^{2}_{\perp}}{4\pi^2 v_x v^{2}_{y}\Lambda} \times
 \begin{cases}
 1 & \text{if } t_{\perp}\text{ is }\bm{k}-\text{independent} \\
 \tau^{2}_{1}(\Q)       &\text{if } t_{\perp}\text{ is } \bm{k}-\text{dependent}
  \end{cases}}
\end{equation}
where the velocities are defined at the hotspots, so depend implicitly on $\Q$. {\it We therefore see that the bilinear coefficient $\alpha$ is strongly suppressed by an interlayer tunneling that is momentum dependent}.

\subsection{Evaluation of $\beta$}
We have $\beta = \beta_1 + \beta_2 + 2\beta_3$, where
\begin{align}
\beta_1 &= 2\int_{\k,\omega_n} G^4_1G^2_2 + G^4_3G^2_4 =2\left( I_3 + I_4\right)\\
\beta_2 &=2\int_{\k,\omega_n} G^4_2G^2_1 + G^4_4G^2_3 =2\left(I_5 + I_6\right)\\
\beta_3 &= 4\int_{\k,\omega_n} G^4_1G_2G_{1-x} + G^4_3G_4G_{3-x}=4\left(I_7 + I_8\right).
\end{align}

\subsubsection{$\beta_1$ Integral}
For $I_3$ we have 
\begin{align}
I_3 = \frac{T}{4\pi^2} \sum_{i\omega_n} \int^{\Lambda}_{-\Lambda} dk_x \int^{\Lambda}_{-\Lambda} dk_y \left[\frac{1}{i\omega+ v_xk_x - v_yk_y}\right]^4\left[\frac{1}{i\omega - v_xk_x - v_yk_y}\right]^2.
\end{align}
We again introduce $x=v_xk_x$, $y=v_yk_y$ and $\Lambda_x$ and $\Lambda_y$ as before, with $\Lambdax \gg \Lambday$. The integral becomes   
\begin{align}
I_3 = \frac{T}{4\pi^2v_xv_y} \sum_{i\omega_n} \int^{\Lambda_y}_{-\Lambda_y} dy \int^{\Lambda_x}_{-\Lambda_x} dx \left(\frac{1}{x-y+i\omega_n}\right)^4\left(\frac{1}{x+y-i\omega_n}\right)^2.
\end{align}
Once more, set $\int^{\Lambdax}_{-\Lambdax} dx = \int^{\infty}_{-\infty}dx - \int_{|x|>\Lambda_x} dx = I_{3a} - I_{3b}$, so that 
\begin{equation}
\begin{aligned}
I_{3a}  &= \frac{T}{4\pi^2v_xv_y} \sum_{i\omega_n} \int^{\Lambda_y}_{-\Lambda_y} dy \int^{\infty}_{-\infty} dx \left(\frac{1}{x-y+i\omega_n}\right)^4\left(\frac{1}{x+y-i\omega_n}\right)^2\\
&\approx \frac{iT}{128\pi^2 v_x v_y} \int d\omega \left[\left(\frac{1}{|\omega|-i\Lambda_y}\right)^4 -  \left(\frac{1}{|\omega| + i\Lambda_y}\right)^4\right]\\
&=\frac{1}{96\pi^2v_xv_y}\frac{1}{\Lambda^{3}_{y}}.
\end{aligned}
\end{equation}
For $I_{3b}$ we follow a similar sequence of procedures as for $I_{1b}$: i.e. change the order of integration and using the assumption that $\Lambda_y \ll \Lambda_x$ perform the $y$ integral first. We get 
\begin{equation}
\begin{aligned}
I_{3b}&=\frac{T}{4\pi^2v_xv_y} \sum_{i\omega_n} \int_{|x|>\Lambda_x} \int^{\Lambday}_{-\Lambday}  \left(\frac{1}{x-y+i\omega_n}\right)^4\left(\frac{1}{x+y-i\omega_n}\right)^2 \\
&\approx \frac{\Lambda_{y}}{2\pi^3 v_x v_y} \int d\omega \int^{\infty}_{\Lambda_x} dx \frac{x^2 - \omega^2}{\left(x^2 + \omega^2\right)^4}\\
&=\frac{1}{32\pi^2 v_x v_y}\frac{\Lambda_y}{\Lambda^{4}_{x}}.
\end{aligned}
\end{equation}
Thus we find that 
\begin{align}
I_{3} &= I_{3a} - I_{3b}\nonumber\\
&=\frac{1}{96\pi^2v_{x}v_y}\left[\frac{1}{\Lambday^3} - \frac{3\Lambda_y}{\Lambdax^4}\right].
\end{align}
\begin{align}\underline{
\implies I_{3} \approx \frac{1}{96\pi^2 v_x v^{4}_{y}}\frac{1}{\Lambda^3}}
\end{align}

Now for $I_4$, we have 
\begin{equation}
I_4 = \frac{T}{4\pi^2} \sum_{i\omega_n} \int^{\Lambda}_{-\Lambda} dk_x \int^{\Lambda}_{-\Lambda} dk_y \left[\frac{1}{i\omega_n -u_xk_x - u_yk_y}\right]^4\left[\frac{1}{i\omega_n +u_xk_x - u_yk_y}\right]^2,
\end{equation}
and we once more use the same change of variables:
\begin{equation}
x=u_x k_x, \quad y=u_yk_y, \quad \Lambda_x = u_x\Lambda, \quad\text{and    } \Lambda_y = u_y\Lambda_y
\end{equation}
but now $\Lambda_y \gg \Lambda_x$. Under this change of variables, we get 
\begin{equation}
I_4 = \frac{T}{4\pi^2u_xu_y}\int^{\Lambda_x}_{-\Lambda_x} dx \int^{\Lambda_y}_{-\Lambda_y} dy \left(\frac{1}{x+y-i\omega_n}\right)^4 \left(\frac{1}{x-y+i\omega_n}\right)^2
\end{equation}
Since $\Lambda_y \gg \Lambda_x$, we write the $y$ integral as $\int^{\Lambda_y}_{-\Lambda_y}dy = \int^{\infty}_{-\infty} dy - \int_{|y|>\Lambda_y}dy = I_{4a} - I_{4b}$. Yet again we find 
\begin{align} 
I_{4a} = \frac{T}{4\pi^2u_xu_y}\int^{\Lambda_x}_{-\Lambda_x} dx \int^{\infty}_{-\infty} dy \left(\frac{1}{x+y-i\omega_n}\right)^4 \left(\frac{1}{x-y+i\omega_n}\right)^2 = 0.
\end{align}
since the poles are both in the same half plane. $I_{4b}$ also turns out to be zero:
\begin{equation}
\begin{aligned}
I_{4b}&= \frac{T}{4\pi^2u_xu_y}\int^{\Lambda_x}_{-\Lambda_x} dx \int_{|y|>\Lambda_y} dy \left(\frac{1}{x+y-i\omega_n}\right)^4 \left(\frac{1}{x-y+i\omega_n}\right)^2\\
&\approx \frac{\Lambda_x}{4\pi^3 u_x u_y}\int d\omega \int^{\infty}_{\Lambda_y} dy \left[\left(\frac{1}{y-i\omega}\right)^6 -\left(\frac{1}{y+i\omega}\right)^6\right]\\
&=0.
\end{aligned}
\end{equation}
so we find that $I_4=0$
and thus, 
\begin{equation}
\underline{\beta_1 = 2I_3 + 2I_4 \approx \frac{1}{48\pi^2v_xv^{4}_y}\frac{1}{\Lambda^3}}.
\end{equation}
 
 \subsubsection{$\beta_2$ integral}
 We again split the integral $\beta_2$ into integrals evaluated at each of the separate pairs of hotspots. We have $\beta_2 = I_5 + I_6$, with 
 \begin{align}
 I_5&=  \frac{T}{4\pi^2} \sum_{i\omega_n} \int^{\Lambda}_{-\Lambda} dk_x \int^{\Lambda}_{-\Lambda} dk_y \left[\frac{1}{i\omega- v_xk_x - v_yk_y}\right]^4\left[\frac{1}{i\omega + v_xk_x - v_yk_y}\right]^2\\
 I_6&= \frac{T}{4\pi^2} \sum_{i\omega_n} \int^{\Lambda}_{-\Lambda} dk_x \int^{\Lambda}_{-\Lambda} dk_y \left[\frac{1}{i\omega_n +u_xk_x - u_yk_y}\right]^4\left[\frac{1}{i\omega_n -u_xk_x - u_yk_y}\right]^2,
 \end{align}
 and it is easy to check that $I_3 = I_5$ and $I_4=I_6$ due to the $C_4$ symmetry of the band structure.  We therefore (trivially) have 
 \begin{equation}
\underline{\beta_2 = \beta_1 = 2I_5 + 2I_6 \approx \frac{1}{48\pi^2v_xv^{4}_y}\frac{1}{\Lambda^3}}
\end{equation}

\subsubsection{$\beta_3$ integral}
We have $\beta_3 = 4\left(I_7 +I_8\right)$, where $I_7$ is at the first pair of hotspots, while $I_8$ is at the second, with
\begin{equation}
I_7 = \frac{T}{4\pi^2}\sum_{i\omega_n} \int^{\Lambda}_{-\Lambda}dk_x \int^{\Lambda}_{-\Lambda} dk_y \left[\frac{1}{i\omega_n +v_xk_x - v_yk_x}\right]^4\left[\frac{1}{i\omega_n -v_xk_x - v_yk_y}\right]\left[\frac{1}{i\omega_n - \xi_{1-X}}\right].
\end{equation}
Yet again we change to the variables $x$, $y$ with $\Lambda_x \gg \Lambda_y$ which yields:
\begin{equation}
I_7 = \frac{T}{4\pi^2v_xv_y}\sum_{i\omega_n}\int^{\Lambda_y}_{-\Lambday} dy \int^{\Lambda_x}_{-\Lambda_x} dx \left(\frac{1}{x-y+i\omega_n}\right)^4\left(\frac{1}{x+y-i\omega_n}\right)\left(\frac{1}{\xi_{1-X} - i\omega_n}\right).
\end{equation}
And we again separate the $x$ integral as follows: $I_{6} = \int^{\infty}_{-\infty} dx - \int_{|x|>\Lambdax}dx$, so that 
\begin{equation}
\begin{aligned}
I_{7a} &= \frac{T}{4\pi^2v_xv_y}\sum_{i\omega_n}\int^{\Lambda_y}_{-\Lambday} dy \int^{\infty}_{-\infty} dx \left(\frac{1}{x-y+i\omega_n}\right)^4\left(\frac{1}{x+y-i\omega_n}\right)\left(\frac{1}{\xi_{1-X} - i\omega_n}\right)\\
&=\frac{1}{192\pi^2 v_xv_y}\int d\omega \,\,\text{sgn}{(\omega)}\left[\left(\frac{1}{\omega-i\Lambda_y}\right)^3 -  \left(\frac{1}{\omega+i\Lambda_y}\right)^3\right] \left(\frac{1}{\xi_{1-X} - i\omega}\right)\\
&=\frac{1}{96\pi^2 v_xv_y}\left[\frac{\varepsilon _0^2 \Lambda _y^2 \left(1-3 \log \left(\frac{\xi _{1-X}}{\Lambda _y}\right)\right)-\Lambda _y^4 \left(2 \log \left(\frac{\xi_{1-X}}{\Lambda _y}\right)+3\right)+\varepsilon _0^4}{\Lambda _y \left(\Lambda _y^2-\xi _{1-X}^2\right){}^3}\right].
\end{aligned}
\end{equation}
We can assume $\Lambda_y \ll \xi_{1-X}$ (i.e. the cutoff of the linearized dispersion is smaller than the `off-shell' dispersion), and expand this expression in terms of $\Lambda_y/\xi_{1-X}$, which gives:
\begin{align}
I_{7a} = -\frac{1}{96\pi^2v_xv^{}_{y}}\left(\frac{1}{v^3_y\Lambda^3}\right)\left[\left( \frac{v_y\Lambda}{\xi_{1-X}}\right)^2 + \left(\frac{v_y\Lambda}{\xi_{1-X}}\right)^4\left( 5 +  6\log{\left(\frac{v_y\Lambda}{\xi_{1-X}}\right)}\right)+ \ldots \right].
\end{align}
So to lowest order, we find 
\begin{align}
I_{7a} \approx -\frac{1}{96\pi^2v_xv^{2}_y}\frac{1}{\Lambda\xi^{2}_{1-X}}.
\end{align}
The integral $I_{7b}$ follows the usual sequence of manipulations. We have
\begin{equation}
\begin{aligned}
I_{7b}&=\frac{T}{4\pi^2v_xv_y}\sum_{i\omega_n}\int_{|x|>\Lambda_x} \int^{\Lambda_y}_{-\Lambda_y} dy \left(\frac{1}{x-y+i\omega_n}\right)^4\left(\frac{1}{x+y-i\omega}\right)\left(\frac{1}{\xi_{1-X} - i\omega_n}\right)\\
&\approx \frac{T\Lambda_y}{2\pi^2v_xv_y} \sum_{i\omega_n} \int_{|x|>\Lambda_x}dx \left(\frac{1}{x+i\omega_n}\right)^4\left(\frac{1}{x-i\omega_n}\right)\left(\frac{1}{\xi_{1-X} - i\omega_n}\right)\\
&\approx \frac{\Lambda_y}{4\pi^3v_xv_y} \int d\omega \int^{\infty}_{\Lambda_x} dx \left(\frac{1}{x^2 + \omega^{2}_{n}}\right)\left[\left(\frac{1}{x+i\omega}\right)^3 - \left(\frac{1}{x-i\omega}\right)^3\right]\left(\frac{1}{\xi_{1-X} - i\omega}\right)\\
&= \frac{\Lambda_y}{4\pi^2v_xv_y}\frac{1}{24 \xi_{1-X} ^3} \left[ \left(-\frac{8 \xi_{1-X} ^2}{(\Lambda_x +\xi_{1-X} )^3}+\frac{3 \xi_{1-X} }{\Lambda_x ^2}-\frac{6 \xi_{1-X} }{(\Lambda_x +\xi_{1-X} )^2}-\frac{6}{\Lambda_x +\xi_{1-X} }\right)+\frac{6}{\xi_{1-X}} \log \left(\frac{\Lambda_x +\xi_{1-X} }{\Lambda_x }\right)\right]\\
&\approx  \frac{1}{32\pi^2v_xv_y}\frac{\Lambda_y}{\Lambda_x^2\xi^{2}_{1-X}} + \ldots
\end{aligned}
\end{equation}
So we find that to lowest order, the integral $I_7$ is given by 
\begin{equation}
\begin{aligned}
I_{7}&= I_{7a}-I_{7b}\\
&=\frac{1}{96\pi^2v_xv_y}\left[-\frac{1}{v_y\Lambda\xi^{2}_{1-X}} -\frac{3v_y\Lambda}{2v^{2}_x\Lambda^2\xi^{2}_{1-X}}\right],
\end{aligned}
\end{equation}
\begin{equation}
\implies\underline{ I_{7}\approx -\frac{1}{96\pi^2v_xv^{2}_y}\frac{1}{\Lambda\xi^{2}_{1-X}}}.
\end{equation}
So we see that $I_7$ is suppressed relative to $I_5$ by a power of $(v_{y}\Lambda/\xi_{1-X})^2$. 
Finally we will evaluate $I_8$, which at the hotspots with velocity $\vec{u}$, with $u_y \gg u_x$, we have 
\begin{equation}
I_8 = \frac{T}{4\pi^2}\sum_{i\omega_n} \int^{\Lambda}_{-\Lambda}dk_x \int^{\Lambda}_{-\Lambda} dk_y \left[\frac{1}{i\omega_n +u_xk_x - u_yk_x}\right]^4\left[\frac{1}{i\omega_n -u_xk_x - u_yk_y}\right]\left[\frac{1}{i\omega_n - \xi_{3-X}}\right].
\end{equation}
Since $\Lambda_y \gg \Lambda_x$, we do the same substitutions as before and split this integral into $I_{8} = \int^{-\Lambda_y}_{\Lambda_y} dy = \int^{\infty}_{-\infty}dy - \int_{|y|>\Lambday}dy = I_{8a}- I_{8b}$ The integral $I_{8a}$ once more vanishes because the poles are both in the same half plane:
\begin{align}
I_{8a} = \frac{T}{4\pi^2u_xu_y} \sum_{i\omega_n}\int^{\Lambda_x}_{-\Lambda_x}dx \int^{\infty}_{-\infty}dy\left(\frac{1}{x-y+i\omega_n}\right)^4\left(\frac{1}{x+y-i\omega_n}\right)\left(\frac{1}{\xi_{3-X} - i\omega_n}\right) =0. 
\end{align}
This time however, the integral $I_{8b}$ does not vanish. We find 
\begin{equation}
\begin{aligned}
I_{8b} &= \frac{T}{4\pi^2u_xu_y} \sum_{i\omega_n} \int^{}_{|y|>\Lambday} dy \int^{\Lambda_x}_{-\Lambda_x}dx\left(\frac{1}{x-y+i\omega_n}\right)^4\left(\frac{1}{x+y-i\omega_n}\right)\left(\frac{1}{\xi_{3-X} - i\omega_n}\right)\\
&\approx \frac{T\Lambda_x}{2\pi^2u_xu_y} \sum_{i\omega_n} \int^{}_{|y|>\Lambday} dy \left(\frac{1}{y-i\omega_n}\right)^5\left(\frac{1}{\xi_{3-X} - i\omega_n}\right)\\
&\approx \frac{\Lambda_x}{4\pi^3u_xu_y}\int d\omega \int^{\infty}_{\Lambday} \left[ \left(\frac{1}{y-i\omega}\right)^5 - \left(\frac{1}{y+i\omega}\right)^5\right]\left(\frac{1}{\xi_{3-X} - i\omega}\right)\\
&=\frac{\Lambda_x}{4\pi^3u_xu_y}\left(\frac{-\pi}{2(\xi_{3-X} +\Lambda_y)^4}\right)\\
&\approx  -\frac{1}{8\pi^2u_xu_y}\frac{u_x\Lambda}{\xi^{4}_{3-X}},
\end{aligned}
\end{equation}
and so we find that 
\begin{equation}
\underline{
I_{8}\approx  \frac{1}{8\pi^2u_xu_y}\frac{u_x\Lambda}{\xi^{4}_{3-X}}
}
\end{equation}

\vspace{5mm}

\subsubsection{Final expression for $\beta$}
So we can now write down the full expression for $\beta$. We have found that 
\begin{equation}
\begin{aligned}
\beta &= \beta_1 + \beta_2 + 2\beta_3\\
&=2\beta_1 + 2\beta_3\\
&= 2(2I_3 + 2I_4) + 2(4I_7 + 4I_8)
\end{aligned}
\end{equation}
Substituting our above results for $I_3$ through $I_8$ we get 
\begin{equation}
\boxed{
\beta \approx \frac{1}{24\pi^2v_xv_y}\frac{1}{v^{3}_y\Lambda^{3}}\left[ 1 - 2\left(\frac{ v_y\Lambda}{\xi_{1-X}}\right)^2\right] + \frac{1}{\pi^2u_xu_y}\frac{1}{\left(u_x\Lambda\right)^3}\left(\frac{u_x\Lambda}{\xi_{3-X}}\right)^4
}
\end{equation}
where one should recall that $v_y\ll v_x$ and $u_x\ll u_y$, while $|\xi_{1-X}|\gg v_y\Lambda$ and $|\xi_{3-X}|\gg u_x\Lambda$.

\subsection{Evaluation of $\gamma$}
The analytic expression for $\gamma$ is (in terms of hotspot integrals) 
\begin{align}
\gamma &= 8\int_{\k,\omega_n} G^{4}_{1}G_{2}\left[G_{1+y} + G_{1-y}\right] + G^{4}_{3}G_{4}\left[G_{3+y} + G_{3-y}\right]\nonumber\\
&= 8\left[ I_9 + I_{10} + I_{11} + I_{12} \right],
\end{align}
The integral $I_9$ is 
\begin{align}
I_9 = \frac{T}{4\pi^2}\sum_{i\omega_n} \int^{\Lambda}_{-\Lambda}dk_x \int^{\Lambda}_{-\Lambda} dk_y \left[\frac{1}{i\omega_n +v_xk_x - v_yk_x}\right]^4\left[\frac{1}{i\omega_n -v_xk_x - v_yk_y}\right]\left[\frac{1}{i\omega_n - \xi_{3-X}}\right].
\end{align}
so we see that this is the same as the integral $I_7$ but with $\xi_{1-X} \rightarrow \xi_{1+Q_x}$. Thus we merely quote the above result:
\begin{equation}
\underline{ I_{9}\approx -\frac{1}{96\pi^2v_xv^{2}_y}\frac{1}{\Lambda\xi^2_{1+Y}}}.
\end{equation}
By similar considerations, we see that $I_{10}$ is also related to this $I_7$, but with $\xi_{1-X} \rightarrow \xi_{1-y}$. So we have 
\begin{equation}
\underline{ I_{10}\approx -\frac{1}{96\pi^2v_xv^{2}_y}\frac{1}{\Lambda\xi^2_{3+Y}}}.
\end{equation}
Identical operations show that $I_{11} = I_8$ with a different constant of the dispersion ($\xi_{3-X} \rightarrow \xi_{3+y}$), and $I_{12} = I_8$ with $\xi_{3-X} \rightarrow \xi_{3-y}$. So we can write 
\begin{equation}
\underline{
I_{11}\approx  \frac{1}{8\pi^2u_xu_y}\frac{u_x\Lambda}{\xi^2_{3+Y}}
},
\end{equation}
and 
\begin{equation}
\underline{
I_{12}\approx  \frac{1}{8\pi^2u_xu_y}\frac{u_x\Lambda}{\xi^2_{3-Y}}
}.
\end{equation}
So we find that 
\begin{equation}
\boxed{
\gamma \approx  -\frac{1}{12\pi^2 v_x v_y}\frac{1}{v^3_y \Lambda^3}\left[\left(\frac{v_y\Lambda}{\varepsilon_{1+Y}}\right)^2 +\left(\frac{v_y\Lambda}{\xi_{1-Y}}\right)^2 \right] + \frac{1}{\pi^2u_xu_y}\frac{1}{\left(u_x\Lambda\right)^3}\left[ \left(\frac{u_x\Lambda}{\xi_{3+Y}}\right)^4 +  \left(\frac{u_x\Lambda}{\xi_{3-Y}}\right)^4\right]
}
\end{equation}

%\end{widetext}
\end{document}